\documentstyle[eqsecnum,prd,epsf,aps]{revtex}

\draft
\begin{document}
%\newcommand{\gsim}
%   {\mathrel{\raise.3ex\hbox{$>$\kern-.75em\lower1ex\hbox{$\sim$}}}}
%\newcommand{\lsim}
%   {\mathrel{\raise.3ex\hbox{$<$\kern-.75em\lower1ex\hbox{$\sim$}}}}

\thispagestyle{empty}
{\baselineskip0pt
\leftline{\large\baselineskip16pt\sl\vbox to0pt{\hbox{\it Department of
      Physics, Kyoto University}
               \hbox{\it Yukawa Institute for Theoretical Physics}\vss}}
\rightline{\large\baselineskip16pt\rm\vbox to20pt{\hbox{KUNS 1405}
             \hbox{YITP/96-31}
%           \hbox{June 1996}
               \hbox{\today}
\vss}}%
}
\vskip1cm
%\vfill
\begin{center}{\large \bf
Scalar gravitational wave from Oppenheimer-Snyder collapse in
scalar-tensor theories of gravity
}
\end{center}
\vskip1cm
\begin{center}
 {\large 
Tomohiro Harada$^{1}$
\footnote{ e-mail: harada@tap.scphys.kyoto-u.ac.jp} } \\
{\large Takeshi Chiba$^{2}$
\footnote{ e-mail: chiba@yukawa.kyoto-u.ac.jp}}\\
{\large Ken-ichi Nakao$^{1}$
\footnote{ e-mail: nakao@tap.scphys.kyoto-u.ac.jp}}\\
{\large Takashi Nakamura$^{2}$
\footnote{ e-mail: takashi@yukawa.kyoto-u.ac.jp}}\\
{\em $^{1}$Department of Physics,~Kyoto University,} 
{\em Kyoto 606-01,~Japan}\\
{\em $^{2}$Yukawa Institute for Theoretical Physics,~Kyoto University,}
{\em Kyoto 606-01,~Japan}\\
\end{center}
%\vfill

\begin{abstract}

Unlike general relativity, scalar-tensor theories of gravity 
predict scalar
gravitational waves even from a spherically symmetric gravitational 
collapse.
We solve numerically the generation and propagation
of the scalar gravitational wave from a spherically
symmetric and homogeneous dust collapse under the approximation
that we can neglect the back reaction of the scalar wave
on the space-time, and examine how the amplitude, characteristic
frequency, and wave form of the observed scalar gravitational wave
depend on the initial radius and mass of the dust and parameters 
contained in the theory.
In the Brans-Dicke theory, through the observation of the scalar gravitational
wave, it is possible to determine the initial radius and mass
and a parameter contained in the theory.
In the scalar-tensor theories, it would be possible to get the information
the first derivative of the coupling function contained in the theory 
because the wave form of the scalar
gravitational wave greatly depends on it.  

\end{abstract}

%\vfill
%\newpage

%******************************************************%
%                     Chapter 1                        %
%******************************************************%

\section{INTRODUCTION}

Until now general relativity has passed all the test experiments
in the solar system and pulsar-timing tests with high accuracy.
However, general relativity is not the only gravitational
theory that passes such tests.
There are alternative theories of gravity that pass 
the present weak-field gravitational tests.  
In the strong-field regime, these theories may deviate  very 
widely 
from general relativity.
Therefore experiments in strong-field regimes are necessary in order to 
determine the correct theory of gravity.

In the 1960s scalar-tensor theories of gravity were actively studied as 
alternative theories of gravity.
Recently they have been revived.  
The reasons are that these theories may play an important role
in the hyperextended inflation model~\cite{sa}, 
that these theories arise naturally
as the low-energy limit of string theory~\cite{cfmp,dp}
or other unified theories, 
and that the laser interferometric gravitational wave observatories
(LIGO~\cite{LIGO})
project will be in practical use in the next decade~\cite{thorne} 
and other projects 
(VIRGO~\cite{VIRGO},
GEO~\cite{GEO}, and TAMA~\cite{TAMA}) 
of gravitational wave observations 
are being carried forward
so that high-accuracy tests of 
the scalar-tensor theories may be expected~\cite{will3,snn}.

In the scalar-tensor theories of gravity we consider scalar fields as well as 
a metric tensor as fields that are related to gravity.
The scalar-tensor theories must agree with general relativity
within an accuracy of at least $\sim0.1 \%$ in the weak-field region 
from the results of solar-system
experiments and pulsar tests~\cite{will1}.
But a class of  scalar-tensor theories that satisfies the 
above constraints
may predict order-of-magnitude deviations  from general relativity in 
the regime of strong gravitational field\cite{def1}.
One example of such phenomena in the strong-field regime may be 
the gravitational collapse of a star.
The Brans-Dicke theory~\cite{bd} is the well-known example of 
scalar-tensor theories.
Collapse in the Brans-Dicke theory has been  researched 
in comparison with that in general relativity.
Matsuda and Nariai~\cite{mn} studied a spherically symmetric collapse of
ideal gas by numerical calculations of relativistic 
hydrodynamics in the Brans-Dicke theory.
Shibata, Nakao, and Nakamura~\cite{snn}  
solved numerically a spherically symmetric 
collapse of  inhomogeneous dust and examined the character of the 
gravitational collapse and the wave form of the radiated 
scalar gravitational wave. 
They found that the final state of the collapse is the Schwarzschild
black hole, which is consistent with the theorem of
Hawking~\cite{hawking} that a stationary black hole in 
the Brans-Dicke theory is identical with that of general relativity.
Scheel, Shapiro, and Teukolsky~\cite{sst} found similar results
as Shibata, Nakao, and Nakamura~\cite{snn}.
 
Since in general relativity no gravitational wave is radiated
 from a spherically symmetric collapse,
we cannot obtain any information of collapse
 from observations of gravitational waves if 
realistic gravitational collapses to black hole
are almost spherically symmetric.
However in scalar-tensor theories scalar gravitational waves
are radiated even from spherically symmetric collapse
and these scalar gravitational waves can be detected by
laser interferometric gravitational wave detectors~\cite{snn}.
It is expected that these scalar gravitational waves 
reflect directly not only the stellar 
initial radius, mass and density but also the theory of gravity.

In this paper we take a spherically symmetric, homogeneous dust collapse
as a model of the collapse of the stellar core  and examine the scalar 
gravitational wave radiation and its wave form.
As a first step to understand the nature of the scalar gravitational 
waves in general, we  neglect the back reaction of the scalar
field on the space-time in this paper. 
The method for numerical calculations is similar to that of Cunningham, 
Price, and Moncrief~\cite{cpm}.
We use null coordinates in order to see the propagation of 
the scalar gravitational wave. 

This paper is organized as follows.
In Sec. \ref{sc:st} we review scalar-tensor theories of gravity
and scalar gravitational waves.
In Sec. \ref{sc:be} we explicitly present our approximation and 
show the basic equations used in our numerical
calculation .
In Sec. \ref{sc:nr} we show numerical results and their implications.
Sec. \ref{sc:ds} will be devoted to discussions.
 
We use units of $c=1$.
The Greek indices $\mu,\nu,\cdots$ run over 0,1,2,3.
A comma represents a partial derivative.
We follow Misner, Thorne, and Wheeler's\cite{mtw} sign conventions of the 
metric tensor, 
Riemann tensor, and Einstein tensor.  

%******************************************************%
%                     Chapter 2                        %
%******************************************************%

\section{Scalar-Tensor Theory}
\label{sc:st}

%<<<<<<<<<<<< Section A >>>>>>>>>>>>>%

\subsection{Field Equations}
\label{ss:fieldeqs}

We consider a class of scalar-tensor theories of gravity 
in which gravitation is mediated by a mass-coupled,
long-range scalar field in addition to space-time curvature. 
The action is given by~\cite{will1} 
\begin{equation}
  I=\frac{1}{16\pi}\int\sqrt{-g}\left(\phi R-\frac{\omega(\phi)}{\phi}
  g^{\mu\nu}\phi_{,\mu}\phi_{,\nu}\right)d^4x
  +I_{m}[\Psi_m,g_{\mu\nu}],
\end{equation}
where $g_{\mu\nu}$ is the metric tensor, $R$ is the scalar curvature, and
$\phi$ is the scalar field that couples to gravity. 
$\phi$ approaches  the cosmological
value $\phi_0$ in the asymptotic region.
 $\omega(\phi)$ is a dimensionless arbitrary function of $\phi$.
 $\Psi_m$ represents  matter  
fields, and $I_m$ is 
the action of them.

Varying the action by $g_{\mu\nu}$ and
 $\phi$ we have
\begin{eqnarray}
   R_{\mu\nu}-\frac{1}{2}g_{\mu\nu}R&=&\frac{8\pi}{\phi}T_{\mu\nu}
   +\frac{\omega(\phi)}{\phi^2}\left(\phi_{,\mu}\phi_{,\nu}
   -\frac{1}{2}g_{\mu\nu}g^{\alpha\beta}
   \phi_{,\alpha}\phi_{,\beta}\right) \nonumber \\ 
   & & +\frac{1}{\phi}(\nabla_{\mu}\nabla_{\nu}\phi-g_{\mu\nu}\Box\phi), 
   \label{eq:fe1} \\
   \Box\phi &=& \frac{1}{3+2\omega(\phi)}\left(8\pi T-\frac{d\omega}{d\phi}
   g^{\alpha\beta}\phi_{,\alpha}\phi_{,\beta}\right),
   \label{eq:fe2}
\end{eqnarray}
where $\nabla_{\nu}$ and $\Box$ are a covariant derivative and the 
d'Alembertian of $g_{\mu\nu}$ respectively. 
The equations of motion for matter fields are given by
\begin{equation}
  \nabla_{\nu}T^{\mu\nu}=0,
\end{equation}
where the energy-momentum
tensor $T^{\mu\nu}$ is defined by 
\begin{equation}
  T^{\mu\nu}\equiv \frac{2}{\sqrt{-g}}\frac{\delta I_m[\Psi_m,g_{\mu\nu}]}
  {\delta g_{\mu\nu}}.
\end{equation}

The weak equivalence principle is valid in this theory: 
A test particle which has negligible self-gravitational energy moves
on a geodesic in this frame, which is often called the Brans-Dicke
frame or the Jordan-Fierz frame. 
However, a self-gravitating body moves 
on a different trajectory because of the coupling of the scalar field
to the body's self-gravity and therefore the gravitational
weak equivalence principle is 
not valid (Nordtvedt effect)~\cite{will1,nordtvedt}.

In the asymptotic region, the gravitational ``constant'' $G_0$, 
which is measured in time dilation experiment or 
by observation of the Keplerian
motion of a test particle, is given by~\cite{zag}
\begin{equation}
  G_0=\phi_0^{-1}\frac{4+2\omega(\phi_0)}{3+2\omega(\phi_0)}.
\end{equation}
Taking the  units of $G_0=1$, we have
\begin{equation}
  \phi_0=\frac{4+2\omega(\phi_0)}{3+2\omega(\phi_0)}.
\end{equation}
If $\omega(\phi)$ is a constant, the theory reduces to the Brans-Dicke theory.
In the limit of $\omega\to\infty$, the theory approaches 
general relativity,
 although it does not mean that every solution of the theory approaches
some solution of general relativity in this limit\cite{sst,rb}.

We can take another representation of the theory by the conformal
transformation from the ``physical'' Brans-Dicke frame to the 
``unphysical'' frame in which the Einstein-Hilbert action 
is separated from the scalar field action~\cite{dicke}. 
This new frame is often called the ``Einstein'' frame.
In this frame, the action is~\cite{dn}
\begin{equation}
  I=\frac{1}{16\pi G_*}\int \sqrt{-g_*}(R_* -2g_*^{\mu\nu}\varphi_{,\mu}
  \varphi_{,\nu})d^4 x+I_m[\Psi_m,A^2(\varphi)g_{*\mu\nu}],
\end{equation}
where $G_*$ is the ``bare'' gravitational constant and
 $\varphi$ is the scalar field.
The metric tensor $g_{*\mu\nu}$ in the Einstein frame is related
to the metric tensor $g_{\mu\nu}$ in the Brans-Dicke frame by the
conformal transformation as
\begin{equation}
  g_{\mu\nu}=A^2(\varphi)g_{*\mu\nu}.
\end{equation}
 $R_*$ is the scalar curvature of $g_{*\mu\nu}$. $\varphi$ 
and $A(\varphi)$ 
are determined by
\begin{eqnarray}
  G_* A^2(\varphi) &=& \frac{1}{\phi}, 
  \label{eq:scalartr1}\\
  \alpha^2(\varphi) &=& \frac{1}{3+2\omega(\phi)}, \label{eq:scalartr2}
\end{eqnarray}
where $\alpha(\varphi)$ is defined by
\begin{equation}
  \alpha(\varphi)\equiv\frac{d\ln A(\varphi)}{d\varphi}.
  \label{eq:scalartr3}
\end{equation}
The field equations are given by
\begin{eqnarray}
  & & G_{*\mu\nu} = 
  R_{*\mu\nu}-\frac{1}{2}R_* g_{*\mu\nu} = 8\pi G_* T_{*\mu\nu}
  +2\left(\varphi_{,\mu}\varphi_{,\nu}-\frac{1}{2}g_{*\mu\nu}g_*^{\alpha\beta}
  \varphi_{,\alpha}\varphi_{,\beta}\right), 
  \label{eq:gravity} \\
  & & \Box_* \varphi = -4\pi G_* \alpha(\varphi)T_*,
  \label{eq:scalar}
\end{eqnarray}
with the equations of motion given by
\begin{equation}
  \nabla_{*\nu}T_{*\mu}^{\nu}=\alpha(\varphi)T_*\nabla_{*\mu}\varphi,
\end{equation}
where $\nabla_{*\nu}$ , $\Box_*$ and $R_{*\mu\nu}$ are a covariant derivative
, the d'Alembertian and the Ricci tensor of $g_{*\mu\nu}$, respectively.
 $T_*^{\mu\nu}$ is defined by 
\begin{equation}
  T^{*\mu\nu}\equiv\frac{2}{\sqrt{-g_*}}\frac{\delta I_m[\Psi_m,
  A^2(\varphi)g_{*\mu\nu}]}{\delta g_{*\mu\nu}},
\end{equation}
and is related to the energy-momentum tensor $T_{\mu\nu}$ in the physical 
Brans-Dicke frame by
\begin{equation}
  T_{*\mu\nu}=A^2(\varphi)T_{\mu\nu}.
\end{equation}
$\varphi$ approaches the cosmological value $\varphi_0$ in the 
asymptotic region.
Hereafter we take the units $G_*=1$ .

Generally, $\alpha(\varphi)$ is an arbitrary function that characterizes
the theory. If $\alpha(\varphi)$ is constant, the theory reduces to 
the Brans-Dicke
theory. In the limit $\alpha \to 0$ , the theory approaches  
general relativity.  
We denote 
\begin{eqnarray}
 \alpha_0 &\equiv& \alpha(\varphi_0), \\ 
 \beta_0 &\equiv& \left.\frac{d\alpha(\varphi)}{d\varphi}
 \right|_{\varphi=\varphi_0}.
\end{eqnarray}
In the post-Newtonian approximation, the PPN parameters $\gamma_{Edd}$ and 
 $\beta_{Edd}$ are expressed as~\cite{dn}
\begin{eqnarray} 
  1-\gamma_{Edd}&=&\frac{2\alpha_0^2}{1+\alpha_0^2},
   \label{eq:gamma} \\
  \beta_{Edd}-1&=&  \frac{\beta_0\alpha_0^2}{2(1+\alpha_0^2)^2}.
  \label{eq:beta}
\end{eqnarray}
Results of solar-system test experiments (the time-delay and deflection of
light) and pulsar-timing tests 
constrain $|1-\gamma_{Edd}|$ as\cite{will1,will2} 
\begin{equation}
  |1-\gamma_{Edd}| < 2\times 10^{-3}.
\end{equation}
This inequality is rewritten in terms of $\omega$ and $\alpha_0$ as
\begin{equation}
\omega > 500,~~~\alpha_0^2 < 10^{-3}.
\end{equation}
The deviation of $\beta_{Edd}$ from unity implies the breakdown of the
gravitational weak equivalence principle. Bounds on this effect from 
the results of 
the lunar-laser-ranging experiment 
constrain $|\beta_{Edd}-1|$ as~\cite{dickey}
\begin{equation}
  |\beta_{Edd}-1| < 7\times 10^{-4}.
\end{equation}
This constraint on $\beta_{Edd}$ 
comes only through the combination of $\alpha_0$
and $\beta_0$ in Eq. (\ref{eq:beta}). 
Therefore the smaller 
the $\alpha_0$ is, the looser it constrains
 $\beta_0$. Very recently, Damour and Esposito-Far\`ese~\cite{def2}
found
 that binary-pulsar experiments constrain $\beta_0$ to
\begin{equation}
  \beta_0 > -5,
\end{equation}
independently of the value of $\alpha_0$ 
for a particular choice of the functional form $\alpha(\varphi)$.

%<<<<<<<<<<<< Section B >>>>>>>>>>>>>%
 
\subsection{Scalar Gravitational Wave}
\label{ss:sgw}

We show that a scalar mode as well as tensor modes propagates
as gravitational waves. We consider linear perturbations
from the Minkowskian metric $\eta_{\mu\nu}$ and constant 
scalar field $\phi_0$.
\begin{eqnarray}
  g_{\mu\nu} &=& \eta_{\mu\nu}+h_{\mu\nu}, \\
  \phi &=& \phi_0+\delta \phi.
\end{eqnarray}
Hereafter in this section 
we raise and drop tensor indices by the flat metric
 $\eta^{\mu\nu}$ and $\eta_{\mu\nu}$. 
The field equations (\ref{eq:fe1}) and (\ref{eq:fe2}) 
in the linear perturbation are given by~\cite{bd}
\begin{eqnarray} 
  & &\frac{1}{2}(h_{\mu\;\;,\nu\alpha}^{\;\;\alpha}
  +h_{\nu\;\;,\mu\alpha}^{\;\;\alpha}
  -h_{\mu\nu\;\;\alpha}^{\;\;\;\;,\alpha}-h_{,\mu\nu})
  -\frac{1}{2}\eta_{\mu\nu}(h_{\alpha\beta}^{\;\;\;\;,\alpha\beta}
  -h_{,\alpha}^{\;\;\alpha}) \nonumber \\   
  & & \quad = \frac{8\pi}{\phi_0}T_{\mu\nu} 
  \label{eq:2b1}
  +\frac{1}{\phi_0}(\delta\phi_{,\mu\nu}-\eta_{\mu\nu}\Box\delta\phi), \\
  & & \Box \delta\phi = \frac{8\pi}{3+2\omega_0}T,
  \label{eq:2b2}
\end{eqnarray}
where $\omega_0\equiv\omega(\phi_0)$ and $h\equiv h_{\alpha}^{\;\;\alpha}$.

We introduce
\begin{equation}
  \theta_{\mu\nu}\equiv h_{\mu\nu}+\frac{\delta\phi}
{\phi_0}\eta_{\mu\nu}.
\end{equation}
Then Eqs. (\ref{eq:2b1}) and (\ref{eq:2b2}) become
\begin{eqnarray}
  & & \theta^{\;\;\;\;\;\;\;\;\alpha}_{\mu\alpha,\nu}+
  \theta^{\;\;\;\;\;\;\;\;\alpha}_{\nu\alpha,\mu}-
  \theta_{\mu\nu\;\;\;\;\alpha}^{\;\;\;\;,\alpha}
  -\theta_{,\mu\nu}-
  \eta_{\mu\nu}(\theta^{\;\;\;\;,\alpha\beta}_{\alpha\beta}-
  \theta_{,\beta}^{\;\;\;\;\beta})=\frac{16\pi}{\phi_0}T_{\mu\nu}, \\
  & & \Box\delta\phi=\frac{8\pi}{3+2\omega_0}T,
\end{eqnarray}
where $\theta \equiv \theta_{\alpha}^{\;\;\alpha}$.
We define $\bar{\theta}_{\mu\nu}$ and $\bar{h}_{\mu\nu} $ by
\begin{equation}
  \bar{\theta}_{\mu\nu} \equiv \theta_{\mu\nu}
  -\frac{1}{2}\eta_{\mu\nu}\theta, 
\end{equation}
and
\begin{equation}
  \bar{h}_{\mu\nu} \equiv h_{\mu\nu}
  -\frac{1}{2}\eta_{\mu\nu}h.
\end{equation}
We use the gauge condition defined by
\begin{equation}
  \bar{h}^{\mu\alpha}_{\;\;\;\;,\alpha}
=\frac{\delta\phi^{,\mu}}{\phi_0}.
\end{equation}
Then the field equations are separated as
\begin{eqnarray}
  \Box\bar{\theta}_{\mu\nu} &=& -\frac{16\pi}{\phi_0}T_{\mu\nu}, \\
  \Box\delta\phi &=& \frac{8\pi}{3+2\omega_0}T.
\end{eqnarray}
In the absence of matter these equations are wave equations.
We can use the remaining degrees of gauge freedom to make 
$\bar{\theta}_{\mu\nu}$ transverse-traceless. Then 
the metric perturbations are separated out 
into  the plus mode, cross mode and scalar mode.
The form of the metric perturbation of the plane wave is expressed by
\begin{equation}
  h_{\mu\nu}=
  \pmatrix{
   0 & 0          & 0           & 0 \cr
   0 & h_{+}      & h_{\times}  & 0 \cr
   0 & h_{\times} & -h_{+}      & 0 \cr
   0 & 0          & 0           & 0 \cr
   }
   -
   \frac{\delta\phi}{\phi_0}
   \pmatrix{
   -1 & 0 & 0 & 0 \cr
    0 & 1 & 0 & 0 \cr
    0 & 0 & 1 & 0 \cr
    0 & 0 & 0 & 1 \cr
   },
\end{equation}
where $h_+$ , $h_{\times}$ represent the plus and cross mode, respectively.
We can detect the scalar mode by a laser interferometric 
gravitational wave detector as shown 
by Shibata, Nakao and Nakamura~\cite{snn}.
 If four detectors are available, we can distinguish 
the plus, cross and the scalar modes and
 determine the direction of the  source in principle~\cite{snn}.

 From Eqs. (\ref{eq:scalartr1})-(\ref{eq:scalartr3}), we can relate the scalar 
field perturbation $\delta\phi=\phi-\phi_0$ in the Brans-Dicke 
frame to the scalar field perturbation $\delta\varphi=
\varphi-\varphi_0$ in the Einstein frame 
in the linear approximation as
\begin{equation}
  \delta\phi=-2\alpha_0(1+\alpha_0^2)\delta\varphi.
\end{equation}

%******************************************************%
%                     Chapter 3                        %
%******************************************************%

\section{Basic Equations}
\label{sc:be}
 
%<<<<<<<<<<<< Section A >>>>>>>>>>>>>%

\subsection{$\alpha_0$ Expansion}

Hereafter we adopt the Einstein frame.
We use the Taylor expansion of  $\alpha(\varphi)$ that characterizes 
the coupling of the scalar field and matter as 
\begin{equation}
  \alpha(\varphi)=\alpha_0+\beta(\varphi-\varphi_0)
  +\beta^{(2)}(\varphi-\varphi_0)^2+\cdots ,
\end{equation}
where $\alpha_0$ represents a deviation of the theory 
from general relativity.
We assume that $|\alpha_0|\ll 1$.
Although it may exclude interesting non-perturbative effects,  
such as a {\it spontaneous scalarization}~\cite{def1,def2}, 
we expand the metric tensor, scalar field and energy-momentum
tensor of matters
in terms of $\alpha_0$,
\begin{eqnarray}
  T_{*\alpha\beta}&=&T^{(E)}_{\alpha\beta}+\alpha_0 t^{(1)}
  _{\alpha\beta}+\alpha_0^2 t^{(2)}_{\alpha\beta} + O(\alpha_0^3), \\
  g_{*\alpha\beta}&=&g^{(E)}_{\alpha\beta}+\alpha_0 h^{(1)}
  _{\alpha\beta}+\alpha_0^2 h^{(2)}_{\alpha\beta} + O(\alpha_0^3), \\
  \varphi&=&\varphi_0+\alpha_0 \varphi^{(1)}+\alpha_0^2 \varphi^{(2)}
  +O(\alpha_0^3).
\end{eqnarray}
 From the lowest order of $\alpha_0$ in Eq. (\ref{eq:gravity}), 
we obtain 
\begin{equation}
   G_{\alpha\beta}^{(E)}=8\pi T_{\alpha\beta}^{(E)}.
\end{equation}
This means that $g_{\alpha\beta}^{(E)}$ and $T_{\alpha\beta}^{(E)}$ 
are solutions in general relativity.
 From the next order of $\alpha_0$, we obtain 
\begin{equation}
   G_{\alpha\beta}^{(1)}=8\pi t^{(1)}_{\alpha\beta}.
\end{equation}
This is also the same as the Einstein equations, so that
the metric of the scalar-tensor theory deviates from 
the general relativity by $O(\alpha_0^2)$. 
Therefore we can determine the scalar field up to $O(\alpha_0)$ by 
solving the wave equation of the scalar field as
\begin{equation}
  \Box^{(E)}\varphi=-4\pi\alpha(\varphi)T^{(E)}.
  \label{eq:weeb}
\end{equation}
Back reaction of the scalar field on the space-time  
can be estimated from $O(\alpha_0^2)$ of Eq. (\ref{eq:gravity}) as 
\begin{equation}
   G_{\alpha\beta}^{(2)}=8\pi t_{\alpha\beta}^{(2)}
   +2\left(\varphi^{(1)}_{,\alpha}\varphi_{,\beta}^{(1)}
   -\frac{1}{2}g^{(E)}_{\alpha\beta}
   g^{(E)\mu\nu}\varphi_{,\mu}^{(1)}\varphi_{,\nu}^{(1)}
   \right).
\end{equation}
This means that the back reaction can be neglected if relations 
\begin{equation}
   \varphi^{(1)}\sim 1,
\end{equation}
and
\begin{equation}
   \alpha_0^2|T^{(2)}_{\varphi\alpha\beta}|\ll |T^{(E)}_{\alpha\beta}|,
\end{equation}
are satisfied. Here we have defined the energy-momentum tensor 
$T^{(2)}_{\varphi\alpha\beta}$ of the scalar field as
\begin{equation}
   T^{(2)}_{\varphi\alpha\beta}\equiv\frac{1}{4\pi}\left(
   \varphi^{(1)}_{,\alpha}\varphi^{(1)}_{,\beta}
   -\frac{1}{2}g^{(E)}_{\alpha\beta}
   g^{(E)\mu\nu}\varphi^{(1)}_{,\mu}\varphi^{(1)}_{,\nu}\right).
\end{equation}

%<<<<<<<<<<<< Section B >>>>>>>>>>>>>%

\subsection{Oppenheimer-Snyder Collapse}

We use the Oppenheimer-Snyder model~\cite{os} as the background space-time and 
matter in which the scalar field evolves. The Oppenheimer-Snyder
solution describes a gravitational 
collapse of a homogeneous spherical dust ball.

In the interior of the dust one can write the line element in the closed 
Friedmann form
\begin{eqnarray}
   ds^2 &=& -d\tau ^2+a(\tau)^2(d\chi^2+\sin^2 \chi d\Omega^2) \\
        &=& a(\eta)^2(-d\eta^2+d\chi^2+\sin^2\chi d\Omega^2), \\
   d\Omega^2 &=& d\theta^2+\sin^2\theta d\phi^2,
   \label{eq:inme} 
\end{eqnarray} 
where
\begin{eqnarray}
   a(\eta) &=& \frac{1}{2}a_0 (1+\cos\eta), 
   \label{eq:scalefactor} \\
   \tau(\eta) &=& \frac{1}{2}a_0(\eta+\sin\eta). 
\end{eqnarray}
The density of the dust is given by
\begin{equation}
   \rho(\eta)=\frac{3a_0}{8\pi}a^{-3}=\frac{3}{8\pi {a_0}^2}
   \left\{ \frac{1}{2}(1+\cos \eta) \right\}^{-3}.
   \label{eq:density}
\end{equation}
The ranges of $\eta$ and $\chi$ are
\begin{equation}
   0 \leq \eta < \pi,  
\end{equation}
and
\begin{equation}
   0 \leq \chi \leq \chi_0 < \frac{\pi}{2}.
\end{equation}

Let $r_s(t)$ be the circumferential 
radius of the stellar surface.
Then in the exterior of the dust ($r > r_s(t)$), 
 the space-time is expressed by the 
Schwarzschild metric as
\begin{equation}
   ds^2 = -\left(1-\frac{2M}{r}\right)dt^2 
   +\left(1-\frac{2M}{r}\right)^{-1} dr^2
   +r^2 d\Omega^2. \label{eq:exme}
\end{equation}

The matching conditions of the interior and exterior solutions are that 
the proper circumferential radius of stellar surface 
should agree and that the stellar surface should move on a
geodesic. These matching conditions are expressed as
\begin{eqnarray}
   r_s&=&a(\eta)\sin\chi_0, \\
   M&=&\frac{1}{2}a_0 \sin^3\chi_0,
   \label{eq:mass} \\
   t &=& 2M\ln\left|\frac{\left(\frac{r_{s0}}{2M}-1\right)^\frac{1}{2}
         +\tan\frac{\eta}{2}}
         {\left(\frac{r_{s0}}{2M}-1\right)^\frac{1}{2}
         -\tan\frac{\eta}{2}}
   \right| \nonumber \\
     & & +2M\left(\frac{r_{s0}}{2M}-1\right)^{\frac{1}{2}}
   \left[\eta+\left(\frac{r_{s0}}{4M}\right)
   (\eta+\sin\eta)\right],
\end{eqnarray}
where $r_{s0}\equiv r_s(t=0)$.

%<<<<<<<<<<<< Section C >>>>>>>>>>>>>%

\subsection{Wave Equation}

We rewrite Eq. (\ref{eq:weeb}) under the background of the 
Oppenheimer-Snyder collapse. 
We use the metric in the form (\ref{eq:inme}) and (\ref{eq:exme}).
The wave equation in the interior ( $0\le \chi \le \chi_0$) is given by 
\begin{equation}
   \frac{1}{a^2}\left\{-\frac{1}{a^2}\frac{\partial}{\partial \eta}
        \left(a^2 \frac{\partial \delta\varphi}{\partial \eta}\right)
        +\frac{1}{\sin^2\chi}\frac{\partial}{\partial \chi}
        \left(\sin^2\chi\frac{\partial \delta\varphi}{\partial \chi}
        \right)\right\} 
   = 4\pi\alpha(\varphi)\rho, \label{eq:inwe}
\end{equation}
while  in the exterior ($r>r_s(t)$) it is given by
\begin{equation}
   -\left(1-\frac{2M}{r}\right)^{-1}
   \frac{\partial^2\delta\varphi}{\partial t^2}
       +\frac{1}{r^2}\frac{\partial}{\partial r}\left\{
       r^2\left(1-\frac{2M}{r}\right)
       \frac{\partial \delta\varphi}{\partial r}
       \right\}  \label{eq:exwe}
   = 0.  
\end{equation}
 Here in place of $\delta\varphi$  we define $\zeta$ by
\begin{equation}
   \zeta \equiv \cases{
                a\sin\chi\delta\varphi & (interior) \cr
                r\delta\varphi         & (exterior). \cr
                }
\label{eq:zetadef}
\end{equation}
Substituting Eq. (\ref{eq:zetadef}) 
into Eqs. (\ref{eq:inwe}) and (\ref{eq:exwe}) 
, we have
\begin{eqnarray}
   & & -\frac{\partial^2 \zeta}{\partial \eta^2}
       +\frac{\partial ^2 \zeta}{\partial \chi^2} 
   = -\left( 1+\frac{a^{\prime\prime}}{a} \right)\zeta
       +4\pi \alpha(\varphi)\rho a^3\sin\chi 
       \quad(\mbox{interior}), 
\label{eq:weinetachi}   \\
   & & -\frac{\partial^2 \zeta}{\partial t^2}
       +\frac{\partial ^2 \zeta}{\partial r_{*}^2} 
   = \frac{2M}{r^3}\left(1-\frac{2M}{r}\right)\zeta 
       \quad(\mbox{exterior}),
\label{eq:weextr}
\end{eqnarray}
where $r_*$ is the tortoise coordinate defined by
\begin{equation}
   r_* = r+2M \ln\left(\frac{r}{2M}-1\right),
\end{equation}
and the prime stands for differentiation with respect to $\eta$.
Using the null coordinates, we can rewrite Eqs. (\ref{eq:weinetachi}) 
and (\ref{eq:weextr}) as
\begin{eqnarray}
   & & \frac{\partial^2 \zeta}{\partial u\partial v}   
   = \frac{1}{4}\left(1+\frac{a^{\prime\prime}}{a}\right)\zeta  
   -\frac{3}{8}\alpha(\varphi) 
   a_0\sin\chi 
   \quad (\mbox{interior}), 
   \label{eq:inwenull} \\
   & & \frac{\partial^2 \zeta}{\partial\tilde{u}\partial\tilde{v}}
   =   
   -\frac{M}{2r^3}\left(1-\frac{2M}{r}\right)\zeta
   \quad (\mbox{exterior}),
   \label{eq:exwenull}
\end{eqnarray}
where the  null coordinates are defined by
\begin{eqnarray}
   u &=& \eta-\chi, \\
   v &=& \eta+\chi,
\end{eqnarray}
in the interior and
\begin{eqnarray}
   \tilde{u} &=& t-r_*, \label{eq:uout} \\
   \tilde{v} &=& t+r_*, \label{eq:vout}
\end{eqnarray}
in the exterior,
and we have used Eq. (\ref{eq:density})
to rewrite $\rho$.

The boundary condition at the center of the dust is
 that the derivative 
of the scalar field in the radial direction should be zero, 
i.e.,
\begin{equation}
   \frac{\partial\delta\varphi}{\partial\chi}=0 
   \quad \mbox{at} \quad \chi=0.
\end{equation}

The junction condition on the stellar surface is that $\varphi$ 
and its derivative  in the direction normal to the boundary 
surface should be continuous, i.e.,
\begin{eqnarray}
   \delta\varphi|_{in} &=& \delta\varphi|_{ex},  \\
   n^{\mu}\delta\varphi_{,\mu}|_{in}&=&n^{\mu}\delta\varphi_{,\mu}|_{ex},   
\end{eqnarray}
at $\chi=\chi_0$ (interior) and $r=r_s(t)$ (exterior), where $n^{\mu}$ is the 
normal vector to the boundary surface of the dust.

For simplicity we take the initial condition
that $\varphi=\varphi_0$ and the time derivative
of $\varphi$ vanishes on the initial hypersurface $\eta=0$ and $t=0$. 
Namely in the interior of the dust we set 
\begin{eqnarray}
   \delta\varphi=0, \\
   \frac{\partial\delta\varphi}{\partial\eta}=0,
\end{eqnarray}
at $\eta=0$ , and in the exterior of the dust
\begin{eqnarray}
   \delta\varphi=0, \\
   \frac{\partial\delta\varphi}{\partial t}=0,
\end{eqnarray}
on the first ray $\tilde{u}=\tilde{u}_0\equiv-r_{*s}(t=0)$, 
because $\varphi=\varphi_0$ is the static solution 
in the exterior region that satisfies the boundary condition in the 
asymptotic region.
We can regard this as the initial condition 
of the situation that for $t<0$ a highly relativistic star is in 
equilibrium between the pressure and the gravity, 
and at $t=0$ the pressure is switched off 
and the homogeneous dust begins to 
collapse, because, for highly relativistic matter, the trace of the 
energy-momentum tensor satisfies
\begin{equation}
   T = -\rho +3P =0.
\end{equation}

%******************************************************%
%                     Chapter 3                        %
%******************************************************%

\section{NUMERICAL RESULTS}
\label{sc:nr}

We divide the Oppenheimer-Snyder space-time to three regions 
(A), (B) and (C) as shown in 
Fig. 1 according to Cunningham, Price and Moncrief~\cite{cpm}. 
We use null coordinates in the numerical calculation in 
order to observe propagation of the scalar field 
to the asymptotic region.
The details of our numerical calculation will be described in 
Appendix \ref{sc:numcod}.
As for the coupling function, for simplicity, we assume that
\begin{equation}
\alpha(\varphi)=\alpha_{0}+\beta (\varphi-\varphi_{0}).
\label{eq:alphalinear}
\end{equation}
 
%<<<<<<<<<<<< Section A >>>>>>>>>>>>>%
\subsection{Brans-Dicke Theory}

%<<<<<<<<<<<< Subsection i >>>>>>>>>>>>>%

\subsubsection{Wave Form in the Wave Zone}

We shall see the behavior of the scalar field in the Brans-Dicke 
theory (i.e., $\beta=0$)  to obtain an  
overall picture of the time evolution of the scalar field. 
Fig. 2 shows the wave form of the scalar gravitational
wave by an observer at $r=100M$ 
in the Brans-Dicke theory from the collapse of 
the dust whose initial radius
$r_{s0}$ is $10M$.
The ordinate is $\zeta=r\delta\varphi$.
The solution is 
proportional to the parameter $\alpha_0$
and hereafter we shall normalize $\zeta$ by  $\alpha_0=-0.0316$ 
corresponding to $\omega=500$.
The first ray reaches the present observer at $t\sim 95M$ and then
the scalar field increases from the initial value $\varphi_0$
to the peak.
The amplitude at the peak is estimated as~\cite{snn}
\begin{equation}
   \delta\varphi\sim\frac{\alpha_0 M}{r}.
\end{equation}
After the peak, the value of the scalar field $\varphi$ decreases below
the initial constant asymptotic value $\varphi_0$ 
and increases monotonically toward $\varphi_0$.
A frequency of this last bounce below $\varphi_0$ 
nearly equals 
that of the spherically symmetric, fundamental quasi-normal 
mode~\cite{andersson}.
This result is in a good agreement with Shibata, 
Nakao and Nakamura's 
result of full numerical simulations\cite{snn}.
This also shows that the back reaction of the   
scalar field on the space-time can be neglected 
when $\omega+3/2 \agt 5$, i.e., $|\alpha_0| \alt 0.316$.

%<<<<<<<<<<<< Subsection ii >>>>>>>>>>>>>%

\subsubsection{Time Evolution of the Scalar Field}\label{sss:qs}

Next we shall see the numerical solution $\zeta$ in the interior
of the dust (the region (A)). 
The result is shown in Figs. 3(a-c).
Fig. 3(a) shows the scalar field in the interior of
the dust (the region (A)) in Fig. 1.
The initial dust radius $r_{s0}$ is 10M.
The case of the initial radius $r_{s0}=10M$ is very instructive and 
therefore we shall first consider this case. 
The abscissas are the null coordinates
$u=\eta-\chi$ and $v=\eta+\chi$.
Fig. 3(b) shows the time evolution of the scalar field
seen by comoving observers.
The abscissa is the conformal time $\eta$ and the numbers attached to 
the curves are the values of the fixed radial coordinates $\chi$ of 
the comoving observers.
Fig. 3(c) shows the time evolution of the scalar field configuration
on the spacelike hypersurface of the conformal time $\eta=const.$.
The abscissa is the radial coordinate $\chi$ and the
numbers attached to the curves are the value of $\eta$. 

As already mentioned, at $\eta=0$ the scalar field is $\varphi_0$, i.e.,  
$\zeta=0$ everywhere.
Then $\varphi$ increases homogeneously in the central region ($u\sim v$) 
because the source of the scalar field is the {\it homogeneous} dust ball and 
the information of the surface of the dust ball has not yet arrived at 
the central region in the early stage. 
The information from the surface of the dust ball propagates inward 
at the speed of light and reaches the center at the time $\eta=\chi_0$. 
After the reflection at the center, the configuration of the scalar field
in the interior of the dust
settles to a {\it quasi-static solution} 
($\partial\zeta/\partial\eta \sim 0$)  
along the outgoing null ray $u\sim\chi_0$ and 
after that the scalar field evolves in a quasi-static manner.
Finally the scalar field falls inside the event horizon, 
keeping its value finite. 

In Sec. \ref{sss:initialradius},
we will see that 
the quasi-static configuration is realized only when 
the initial radius of the dust ball is large enough. 
This fact suggests 
that the quasi-static evolution appears in the case 
$ \chi_{0} \ll 1$, which 
means that the gravity is weak in the initial configuration. 
In order to confirm this expectation, we shall assume that $\chi_{0}\ll 1$ and 
that the interior solution 
$\zeta=\zeta_{in}$ can be expanded with respect to $\chi_{0}$. 
Then we can obtain a consistent quasi-static solution up to
$O(\chi_0^3)$, which is given by
\begin{eqnarray}
\zeta_{in}&=&{1\over4}\alpha_{0}a_{0}\chi(\chi^{2}-3\chi_{0}^{2}), 
\label{eq:qsin} \\
\zeta_{ex}&=&-{1\over2}\alpha_{0}a_{0}\chi_{0}^{3}\sim -\alpha_{0}M.
\label{eq:qsex}
\end{eqnarray}
The derivation of this solution will be described 
in Appendix \ref{sc:qssol}.     
We plot the interior solution (\ref{eq:qsin}) 
in Fig. 3(c) with the label ``QS''.
We see 
that the approximate solution agrees with the numerical solutions very well 
for the case of $\beta=0$ even when the dust surface is rather 
close to the event horizon. 
When the surface of the dust ball is close to the event horizon, 
the amplitude of the scalar field in the interior region is 
\begin{equation}
   \delta\varphi \sim -\frac{\alpha_0 M}{r_s},
\end{equation} 
and the ratio of the energy density of the scalar field to that of the dust 
remains $O(\alpha_0^2)$.

The numerical solution $\zeta=r\delta\varphi$ in the exterior of the dust
is shown in Figs. 4 (the region (B)) and 5 (the region (C)).
The set of null coordinates $u$ , $v$ of Fig. 4 is 
a continuous extension of the interior characteristic coordinates
obtained by relabeling the rays $\tilde{u}=const.$ and $\tilde{v}=const.$. 
The details of this coordinate extension will be described in 
Appendix \ref{sc:coordext}.
In the region (B) the scalar field increases and settles 
to the quasi-static configuration on the null ray $u\sim \chi_0$.
In the region (C) the outgoing radiation of the scalar field can be seen. 
If we retrieve $G$ and $c$, the units of time and length become
\begin{eqnarray}
   \frac{GM}{c^3} &=& 4.93\times10^{-6}\frac{M}{M_{\odot}}\quad\mbox{sec}, \\
   \frac{GM}{c^2} &=& 1.48\frac{M}{M_{\odot}} \quad \mbox{km},
\end{eqnarray}
respectively.

As seen in Figs. 4 and 5, first the scalar field increases 
from the initial value $\varphi_0$ due to the presence of the dust.
When the event horizon is formed, the field inside cannot affect the field 
outside. At that moment no wave propagates outwardly from the region (B) to the
region (C). In the wave zone the scalar field approaches 
the asymptotic value $\varphi_0$ 
after the wave has passed 
the observer at $r=const.$.
This is consistent with the Hawking's 
theorem which states that the black hole 
has no scalar charge~\cite{hawking}. 
In Fig. 5 a steep cliff
can be seen on the boundary $\tilde{v}=\tilde{v_0}$ between 
the regions (B) and (C) but this is only due to the coordinate singularity
of the Schwarzschild coordinates and does not mean any singular
behavior of the scalar field. 

%<<<<<<<<<<<< Subsection iii >>>>>>>>>>>>>%
\subsubsection{Dependence on the Initial Radius of the Dust}
\label{sss:initialradius}

In Fig. 6, wave forms of the scalar gravitational waves
from the collapse of the dust whose initial radii $r_{s0}$ are 
$3M$, $4M$, $6M$, $8M$, $10M$ and $20M$ in the Brans-Dicke theory 
are plotted. 
As we can see in Fig. 6, peaks
are nearly symmetric for $r_{s0}\alt 4M$ but not for $r_{s0}\agt 4M$.
Fig. 7 shows the relation between the peak amplitude of the wave form
and the initial dust radius.
As seen from Fig. 7
, the amplitudes at the peaks are 
nearly $|\alpha_0| M/r$ for $r_{s0}\agt 4M$ 
but smaller than $|\alpha_0| M/r$  
for $r_{s0}\alt 4M$. These can be explained as follows.
The information on the stellar surface reaches the center of the dust
first at $\eta=\chi_0$. On the other hand the information
at the center at $\eta\geq\pi-3\chi_0$ cannot travel to the exterior
because it is inside the event horizon.
Therefore for the dust of $\chi_0\agt \pi/4$, i.e.,
$r_{s0}\alt 4M$, the scalar field cannot reach the quasi-static
evolution until the event horizon is formed.
So a mild downhill after the peak, which is an evidence of the 
quasi-static evolution in the interior of the dust
, cannot be seen for $r_{s0}\alt 4M$.
On the other hand, for $r_{s0}\agt 4M$, a mild
down hill after the peak implies the quasi-static evolution
in the interior of the dust.

For $ r_{s0}\agt 4M$,
since the initial ingoing null ray from the
stellar surface bounces back to the surface  $\eta\sim 2\chi_0$, 
the time $\Delta t_{peak}$ that it takes for the scalar 
field to reach the peak
from its rise is estimated as
\begin{eqnarray}
   \Delta t_{peak} &\sim& t(\eta=2\chi_0) \nonumber \\
                   &=& 2M\ln \left| \frac{\cot\chi_0+\tan\chi_0}
                       {\cot\chi_0-\tan\chi_0}\right| \nonumber \\
                   &+& 2M\cot\chi_0\left[2\chi_0+\frac{1}{2\sin^2\chi_0}
                        (2\chi_0+\sin 2\chi_0)\right] .
\label{eq:tpeak}
\end{eqnarray}
This shows an agreement with the numerical results in Fig. 6
within an accuracy of $\sim$ 10 \%.
For example, using Eq. (\ref{eq:tpeak}), we can 
estimate $\Delta t_{peak}$ to be $\sim 40 M$ for
$r_{s0}=20M$.

Fig. 8 shows the relation between an initial dust radius $r_{s0}$
and a characteristic frequency $f_c$ which is defined as the
frequency at which the energy spectrum of the wave form
takes the maximal value.
For $r_{s0}\agt 8M$, the characteristic frequency $f_c$ is 
approximately in proportion to
the inverse of the free-fall time of the initial dust, which is shown by
Shibata, Nakao and Nakamura~\cite{snn} as
\begin{equation}
   f_c\propto \frac{1}{t_{ff}} \sim \sqrt{\rho_0}.
\end{equation} 
For $r_{s0}\alt 8M$, though there is no such simple relation  
because of the effect of  the space-time curvature, $f_c$ would be 
determined by the initial radius and mass of the dust.
The last bounce below the asymptotic value $\varphi_0$ 
reflects a quasi-normal
oscillation.
Fig. 9 shows the relation between the initial dust radius and
the time scale of the bounce, which is defined
as the half value width of that bounce.
As shown in Fig. 9, the time scale of this 
bounce is almost independent from the initial dust radius
and nearly equals to the period of 
the quasi-normal oscillation~\cite{andersson}.

%<<<<<<<<<<<< Section B >>>>>>>>>>>>>%

\subsection{Dependence of the results on the Parameter in the Theory}
 
In Figs. 10-12 the wave forms of the scalar gravitational waves seen
by an observer at $r=100M$ are shown, for various $\beta$, 
from the collapse of the dust whose initial
radius is fixed to $10M$.
In each figure, the scalar field approaches the asymptotic value after the 
peak, the quasi-normal mode of the scalar wave having passed the point of
the observer, which is consistent with the 
black hole no hair conjecture~\cite{rw}.

First we consider the case of positive $\beta$. The amplitude of the 
scalar gravitational wave decreases as $\beta$ increases.
As seen from Figs. 10 and 11 
there appears another oscillation mode whose frequency is characteristic
of the relevant gravitational theory and the dust density.
The larger $\beta$ is, the shorter is the period of this 
oscillation mode  and the more the scalar field oscillates 
before it approaches the asymptotic value. 
This can be also seen from Fig. 13, 
which shows the time of the first minimum from the
reach of the first ray.
This oscillation mode reflects an oscillation mode 
in the interior of the dust seen
in Fig. 14,
which shows the time evolution of the scalar field observed by 
comoving observers in the region(A)
for $\beta=50$ .
The ratio of the energy density of the scalar field to that of the dust
remains within $\sim O(\alpha_0^2)$. 
So, if $|\alpha_0|\ll 1$, the back reaction of the scalar field on the
space-time is negligible.

Second we consider the case of negative $\beta$.
The amplitude of the scalar wave increases
as $|\beta|$ increases, as seen in Fig. 12.
This reflects an exponential growth 
of the scalar field in the interior 
of the dust seen in Fig. 15, 
which shows the time evolution of the scalar field observed 
by comoving observers in the region(A) for $\beta=-5$ . 
The ratio of the energy of the scalar field to that of the
dust is $\sim 10\times O(\alpha_0^2)$ for $\beta=-1$, $\sim 100\times
 O(\alpha_0^2)$ for $\beta=-5$ and $\sim 10^4\times O(\alpha_0^2)$ for
$\beta=-10$. Therefore, 
if we consider a finite value of $\alpha_0$,
there is a critical value of $\beta$ for which the back reaction of the
scalar field cannot be neglected.

As we have seen in Sec. \ref{ss:fieldeqs}, in fact,
the solar-system test experiments for the post-Newtonian order
deviation from general relativity do not constrain directly the value 
of $\beta$ because $\beta$ is constrained only through the
combination with $\alpha_0$ in Eq. (\ref{eq:beta}) (however, see \cite{def2}). 
So the wave form for $\beta$, the value of which is allowed 
by the solar-system experiments, may be quite different from
that in the Brans-Dicke theory.

\section{DISCUSSIONS AND SUMMARY}
\label{sc:ds}

In the Brans-Dicke theory the peak amplitude of the scalar
gravitational wave is given by
\begin{equation}
   \delta\phi\sim
   \alpha_0\delta\varphi\sim 10^{-23}
   \left(\frac{500}{\omega}\right)
   \left(\frac{M}{M_{\odot}}\right)
   \left(\frac{10\mbox{Mpc}}{r}\right),
\end{equation}    
and the characteristic frequency is given by 
\begin{equation}
   f_c \sim 3\times10^3 \left(\frac{M}{M_{\odot}}\right)^{1/2}
   \left(\frac{r_{s0}}{15\mbox{km}}\right)^{-3/2}\mbox{Hz},  
\end{equation}
if $r_{s0}\agt 4M$. For this frequency the sensitivity of the first
LIGO would be $h\sim 10^{-21}$ and that of the advanced LIGO would
be $h\sim 10^{-22}$. If $\omega\sim 500$, the advanced LIGO
may detect the scalar gravitational wave from  collapse 
of a nearly spherical mass $\sim 1 M_{\odot}$ and 
initial radius $\sim 15\mbox{km}$ at
the distance $\sim 1 \mbox{Mpc}$ from us.
If the delayed collapse~\cite{brown,bb} after a supernova would happen
in our Galaxy, we can test the Brans-Dicke theory by the
advanced LIGO up to $\omega \alt 10^5$. Even for a neutron star
formation, scalar gravitational wave may be emitted because
the progenitor would lose a considerable part
of its scalar mass in the process.
The amplitude of the wave would tell us  the information of 
its self-gravitational energy.
If a wave form of the scalar gravitational wave from gravitational 
collapse is obtained observationally, 
we can determine its amplitude, 
characteristic frequency and quasi-normal mode frequency.
We can then determine the source mass because
the frequency of the quasi-normal mode is in 
proportion to the inverse of the mass.
If we know, in addition, the distance to
the source by another method, 
we can determine the Brans-Dicke parameter $\omega$.
Moreover, 
we can determine the initial dust radius
from its characteristic frequency. 
In the case of delayed collapse after the supernova, 
the core radius 
would correspond to it.
The information of the initial radius would constrain the possible 
high-density equation of state. 

Next we consider the parameter dependence of the wave form in the 
scalar-tensor theory.
The wave form greatly depends on $\beta$.
If the space-time is flat and the density
is homogeneous,
a homogeneous solution
which describes harmonic oscillations at the period
$\sqrt{\pi/(\beta \rho)}$ for $\beta>0$ and 
an exponential increase at the {\it e}-fold time $1/(2\sqrt{\pi|\beta|\rho})$
for $\beta<0$ appears. 
Such a mode would appear in the wave form of 
the scalar gravitational wave.
For $\beta>0$ the back reaction of the scalar field on the space-time
is negligible, and so we can safely expand the theory 
around general relativity.
Therefore we can determine the wave form of
the scalar gravitational wave within our approximation.
For $\beta<0$ the back reaction is not negligible and 
the full calculation that
deals with the dynamics of the metric, matter and scalar field
is needed in order to predict the correct wave form~\cite{chnn}.
We also see in Eq. (\ref{eq:weeb}) that the scalar field gets
an effective mass $m^2=-4\pi\beta T$.
For $\beta>0$, the scalar field can be regarded as 
being  effectively massive since $T=-\rho+3P<0$ , and
for $\beta<0$, the scalar field gets effectively a {\it tachyonic} mass.

Finally we make a crude estimate of the condition in which 
the scalar mode of gravitational waves 
radiated from its collapse dominates over the 
tensor mode. We consider collapse of an oblate spheroid of mass $M$. 
Then the tensor mode of the wave at distance $r$ takes the form
\begin{equation}
   h_T \simeq {2\over r} \ddot{Q} \simeq {2\over r}M(a^2-c^2)^{\cdot
   \cdot},
\end{equation}
where $a$ and $c$ are the semi
-major axis and semi-minor axis respectively. 
While the scalar mode is
\begin{equation}
   h_S \simeq {1\over \omega}{M\over r},
\end{equation}
where $Q$ is the reduced quadrupole moment and
the dot means a time derivative.
If we replace the time derivative with the free-fall time $t_{ff}$,
then the ratio of $h_T$ to $h_S$ is
\begin{equation}
   h_T/h_S \sim \frac{2\omega}{M}\frac{M(a^2-c^2)}{t_{ff}^2} 
   \sim 16\omega(a^2-c^2)\frac{M}{a^2c} 
   \sim 10^3\left({\omega \over 
   500}\right)\left({10M\over a}\right){e^2\over \sqrt{1-e^2}}.
\end{equation}
For example, for $\omega=500$ and $a=10M$ the scalar mode dominates over
the tensor mode if the eccentricity $e \alt 0.03$. 
Thus the scalar mode would dominate in almost spherical
collapse. 

We have examined a scalar gravitational wave from
a spherically symmetric and homogeneous 
dust collapse in the Brans-Dicke theory and scalar-tensor theories of gravity,
where we have neglected the back reaction of the scalar field
on the space-time.

(1) In the Brans-Dicke theory the back reaction of the scalar 
field on the space-time remains $O(1/\omega)$. 
So if $\omega \gg 1$, it is negligible.

(2) In the Brans-Dicke theory the amplitude
of the scalar gravitational wave is given by $\sim M/(\omega r)$
and the characteristic frequency is given by $\sim 3\times10^3 
(M/M_\odot)^{1/2}
(r_{s0}/15\mbox{km})^{-3/2}\mbox{Hz}$.
The characteristic frequency of the scalar gravitational wave is 
in proportion 
to the inverse of the free-fall time of the initial dust sphere.
The last bounce 
reflects the fundamental quasi-normal mode of the black hole.
If $r_{s0}\agt 4M$, the scalar field in the interior of 
the dust reaches a 
quasi-static configuration and 
thereafter evolves quasi-statically.
Therefore the wave form after the peak goes down slowly.
If $r_{s0}\alt 4M$, the scalar field in the interior of the dust 
does not reach 
the quasi-static configuration
and so the wave form after the peak goes down more quickly.
The peak amplitude of the scalar gravitational wave
is smaller than $M/(\omega r)$ in this case.

(3) Both in the Brans-Dicke theory and in the scalar-tensor theory given 
by Eq. (\ref{eq:alphalinear}), the scalar field approaches
its asymptotic value after the scalar wave radiation 
has passed the observer at $r=const.$.
In our numerical calculation we do not find any singularity in the 
scalar field at the event horizon.

(4) In the Brans-Dicke theory we can determine a mass, initial radius
of the dust and the Brans-Dicke parameter from the wave form
of the scalar gravitational wave and the distance to its source.

(5) We calculate the scalar gravitational wave
from the dust collapse in the scalar-tensor theory
given by Eq. (\ref{eq:alphalinear}).
The wave form greatly depends on the parameter $\beta$.
If $\beta>0$, the amplitude of the scalar gravitational 
wave is suppressed and a new oscillation mode
whose time scale is $\sim 1/\sqrt{\beta\rho}$ appears.
The back reaction of the scalar field on the space-time is 
within $\sim O(\alpha_0^2)$.
If $\beta<0$, the amplitude of the scalar gravitational wave
is enhanced and its back reaction would not be negligible.
Hence we have to solve the whole field equations fully numerically,
which is the subject of our next paper\cite{chnn}.

\acknowledgments
We would like to thank T. Damour for suggesting to investigate
this problem.
We would like to thank M. Siino, M. Shibata, S. Hayward, Y. Fujii 
and H. Sato for useful discussions. 
T.H. and T.C. are also grateful to H.Sato for
continuous encouragement. 
This work was in part supported by the Grant-in-Aid 
for Basic Research (Grant No.08NP0801)
and for Encouragement of Young Scientists (Grant No.08740341)
of Ministry of Education, Culture, Science and Sports.

\appendix

\section{NUMERICAL CODE}
\label{sc:numcod}

The wave equations we calculate take the form.
\begin{eqnarray}
   \frac{\partial q}{\partial u} &=& A(u,v,\xi,p,q) ,\\
   \frac{\partial \xi}{\partial u} &=& p ,\\
   \frac{\partial p}{\partial v} &=& A(u,v,\xi,p,q) ,\\
   \frac{\partial \xi}{\partial v} &=& q ,
\end{eqnarray}
Our numerical code is similar to that of Hamad\'e and Stewart~\cite{hs},
but we slightly modify their finite differential equations.
We solve simultaneous partial differential equations given by
\begin{eqnarray}
   \frac{\partial y}{\partial u} &=& F(y,z), \\
   \frac{\partial z}{\partial v} &=& G(y,z).
\end{eqnarray}
We determine $y_n$, $z_n$ at the point ${\bf n}(u,v)$ when
$y_w$ and $z_w$ are given at the point ${\bf w}(u,v-h)$ and
$y_e$, $z_e$ at the point ${\bf e}(u-h,v)$.
At the first step we calculate
\begin{eqnarray}
   \hat{y_n} &=& y_e + hF(y_e,z_e) \\
   \hat{z_n} &=& z_w + \frac{1}{2}h(G(y_w,z_w)+G(\hat{y_n},\hat{z_n})), 
\end{eqnarray}
and at the second step
\begin{eqnarray}
   y_n &=& \frac{1}{2}(\hat{y_n}+y_e+hF(\hat{y_n},\hat{z_n})) ,\\
   z_n &=& \frac{1}{2}\left\{\hat{z_n}+z_w+\frac{1}{2}h(G(y_w,z_w)
   +G(\hat{y_n},\hat{z_n}))\right\}.     
\end{eqnarray}
In general this is an implicit scheme, but since the right hand side of 
the wave equation
we have calculated 
does not contain any first derivative of the field
as seen Eqs. (\ref{eq:inwenull}), (\ref{eq:exwenull}),
it becomes an explicit scheme.
The error of the one set of these two steps is estimated as $O(h^3)$.

This code was checked through the following tests.
We calculated a spherically symmetric, ingoing wave in flat space-time
which bounces at the center, and compared it with an analytic solution.
The error of this test is within $\sim$ 0.5\%.
We calculated scalar gravitational waves from the Oppenheimer-Snyder
collapse in the Brans-Dicke theory 
and compared them to the results for $\omega=500$ 
of the full simulation by Shibata,
Nakao and Nakamura~\cite{snn}. 
The two results are agreed within an accuracy of several percent.
These tests make us sure that our
numerical code works well. 
It took about two hours to compute one model
on a Sun SPARC Station 20 (50MHz).

\section{COORDINATE EXTENSION}
\label{sc:coordext}

It is convenient for our numerical calculation 
to divide the exterior region into the two
regions (B) and (C) in Fig. 1 by the 
inward light ray
$\tilde{v}=\tilde{v}_0$ which reaches the stellar surface
on the event horizon.
We extend the interior comoving coordinates to the exterior region
following Cunningham, Price and 
Moncrief~\cite{cpm}.
We relabel light rays $\tilde{u}=const.$ and $\tilde{v}=const.$ 
by the interior coordinates $u$, $v$ at which these light rays
reach the stellar surface.
This coordinate transformation is expressed explicitly 
as 
\begin{eqnarray}
   \tilde{u} &=& t_s(u+\chi_0)-r_{*s}(u+\chi_0), \\
   \tilde{v} &=& t_s(v-\chi_0)+r_{*s}(v-\chi_0), \\
   t_s(\eta) &=& 2M\ln\left|\frac{\cot\chi_0+\tan\frac{\eta}{2}}
   {\cot\chi_0-\tan\frac{\eta}{2}}\right| \nonumber \\
   & & +2M\cot\chi_0\left[\eta+\frac{1}{2\sin^2\chi_0}(\eta+\sin\eta)
   \right], \\
   r_{*s}(\eta) &=& a(\eta)\sin\chi_0
         +2M\ln\left|\frac{a(\eta)\sin\chi_0}{2M}-1
   \right|. 
\end{eqnarray}
This new coordinates are well-behaved everywhere in the space-time
except for the central singularity of the space-time.

In these coordinates, the junction conditions on the surface become
particularly simple because the vector normal to the surface is given 
\begin{equation}
   {\bf n}=\frac{1}{a}\left(\frac{\partial}{\partial v}
   -\frac{\partial}{\partial u}\right).
\end{equation}

\section{QUASI-STATIC SOLUTION}
\label{sc:qssol}

We assume that $\chi_{0}\ll 1$ and 
that the interior solution 
$\zeta=\zeta_{in}$ can be expanded with respect to $\chi_{0}$ as 
\begin{equation}
\zeta_{in}=\sum_{n=1}^{\infty}\zeta^{(n)}_{in},
\end{equation}
where $\zeta^{(n)}_{in}=O(\chi_{0}^{n})$. 
Since $\chi$ is small, the interior solution 
$\zeta_{in}$ will be well approximated by 
the truncated Taylor series of $\chi$ 
and therefore we should impose 
\begin{equation}
{\partial^{m}\zeta_{in}^{(n)}\over\partial\chi^{m}}=O(\chi_{0}^{n-m}) . 
\end{equation}

Under the assumption of the quasi-static evolution phase, 
$\partial\zeta_{in}/\partial\eta=\partial^{2}\zeta_{in}/\partial\eta^{2}=0$, 
the wave equation (\ref{eq:weinetachi}), in the case of $\beta=0$,
is written up to 
$O(\chi_{0}^{5})$ as 
\begin{eqnarray}
{\partial^{2}\zeta_{in}^{(0)}\over \partial\chi^{2}}&=&0,
\label{eq:order0} \\
{\partial^{2}\zeta_{in}^{(1)}\over \partial\chi^{2}}&=&0,
\label{eq:order1} \\
{\partial^{2}\zeta_{in}^{(2)}\over \partial\chi^{2}}&=&
-{a_{0}\over 2a}\zeta_{in}^{(0)},
\label{eq:order2} \\
{\partial^{2}\zeta_{in}^{(3)}\over \partial\chi^{2}}&=&
-{a_{0}\over 2a}\zeta_{in}^{(1)}
+{3\over2}\alpha_{0}a_{0}\chi, 
\label{eq:order3}\\
{\partial^{2}\zeta_{in}^{(4)}\over \partial\chi^{2}}&=&
-{a_{0}\over 2a}\zeta_{in}^{(2)},
\label{eq:order4} \\
{\partial^{2}\zeta_{in}^{(5)}\over \partial\chi^{2}}&=&
-{a_{0}\over 2a}\zeta_{in}^{(3)}
-{1\over4}\alpha_{0}a_{0}\chi^{3},
\label{eq:order5} 
\end{eqnarray}
where we have used Eqs. (\ref{eq:scalefactor}) 
and (\ref{eq:density}) to rewrite $a^{\prime\prime}$ and 
$\rho$. 

On the other hand, we consider only the 
nontrivial, lowest-order solution 
for the exterior region. 
Since $2M/r_{s}=(a_{0}/a)\sin^{2}\chi_{0}=O(\chi_{0}^{2})$, 
as long as $a_{0}/a$ is not so large, 
the lowest order wave equation for the exterior solution $\zeta=\zeta_{ex}$ 
becomes
\begin{equation}
-{\partial^{2}\zeta_{ex}\over\partial t^{2}}
+{\partial^{2}\zeta_{ex}\over\partial r^{2}}=0.  
\label{eq:weexapp}
\end{equation}
Under the assumption of quasi-static evolution, the matching condition 
between the interior and the exterior solutions at the dust surface 
$r=r_{s}$, that is, $\chi=\chi_{0}$ is given by
\begin{eqnarray}
\zeta_{ex}&=&\zeta_{in}, 
\label{eq:matching1}\\
{\partial\zeta_{ex}\over\partial t}&=&{1\over a}\sin\chi_{0}
\sqrt{\frac{a_{0}}{a}-1}{\partial\zeta_{in}\over \partial\chi}
\sim {1\over a}\sqrt{\frac{a_{0}}{a}-1}
{\partial\zeta_{in}\over \partial\chi}
[\chi_{0}+O(\chi_{0}^{3})],
\label{eq:matching2} \\
{\partial\zeta_{ex}\over\partial r}&=&{1\over a}
\left(1-{a_{0}\over a}\sin^{2}\chi_{0}\right)^{-1}\cos\chi_{0}
{\partial\zeta_{in}\over \partial\chi}
\sim {1\over a}{\partial\zeta_{in}\over \partial\chi}
[1+O(\chi_{0}^{2})].
\label{eq:matching3} 
\end{eqnarray}
 From the above equation, we find that 
\begin{equation}
{|\partial\zeta_{ex}/\partial t|\over|\partial\zeta_{ex}/\partial r|}=
O(\chi_{0})~~~~~{\rm at}~~~~~r=r_{s}.
\end{equation}
This means that the exterior solution satisfies 
$\partial\zeta_{ex}/\partial t=0$ in the lowest order of $\chi_{0}$ 
as long as $a_{0}/a \ll 1/\chi_{0}$ 
and therefore from Eq. (\ref{eq:weexapp}), we obtain
\begin{equation}
\zeta_{ex}=C_{ex}=const.,
\end{equation}
where we have imposed 
the boundary condition $\zeta_{ex}\rightarrow const.$ for 
$r\rightarrow\infty$. 
 From the above equation, the matching condition (\ref{eq:matching3}) 
in the lowest order becomes
\begin{equation}
{\partial\zeta_{in}\over \partial\chi}=0~~~~~~{\rm at}~~~~~~\chi=\chi_{0}.
\label{eq:matching4}
\end{equation}

Now we return to the interior solution. 
The solutions for $\zeta_{in}^{(n)}$ ($n=0,1$) are given by
\begin{equation}
\zeta^{(0)}_{in}=C^{(0)}_{in}~~~~~{\rm and}~~~~~
\zeta^{(1)}_{in}=C^{(1)}_{in}+D^{(1)}_{in}\chi, 
\end{equation}
where $C^{(n)}_{in}$ ($n=0,1$) and $D^{(1)}_{in}$ are 
the integration constants and 
the order of those are, respectively, $C_{in}^{(0)},~D^{(1)}_{in}=O(1)$, 
$C_{in}^{(1)}=O(\chi_{0})$. 
However from the regularity at the origin, $\chi=0$, i.e., 
$\zeta_{in}\propto\chi$ for $\chi\rightarrow 0$, $C_{in}^{(0)}$ 
and $C_{in}^{(1)}$ vanish, while $D^{(1)}_{in}$  is 
determined by the matching condition (\ref{eq:matching4}) 
and we find $D^{(1)}_{in}$ also 
vanishes. 
As a result, we get 
\begin{equation}
\zeta_{in}^{(0)}=0=\zeta_{in}^{(1)},
\end{equation}
Using the above results, we obtain the solutions of the equation for 
$\zeta_{in}^{(n)}$ ($n=2,3$) as 
\begin{equation}
\zeta_{in}^{(2)}=0,~~~~~{\rm and}~~~~~
\zeta^{(3)}_{in}={1\over 4}\alpha_{0}a_{0}\chi^{3}+D^{(3)}_{in}\chi,
\label{eq:2and3}
\end{equation}
where we have used the regularity condition at the origin and $D^{(3)}_{in}$ 
is the integration constant which is determined by the matching condition 
(\ref{eq:matching4}). 
Hence the non-trivial solution for the interior region appears 
at $O(\chi_{0}^{3})$ and using this solution, $C_{ex}$ is determined from 
Eq. (\ref{eq:matching1}). 
Then we obtain the solution given by 
Eqs. (\ref{eq:qsin}) and (\ref{eq:qsex}).
This solution is clearly time independent and is consistent with 
the assumption of quasi-static evolution. 
 From Eq. (\ref{eq:order4}), the first equation of Eq. (\ref{eq:2and3}) 
and the regularity at the origin, 
we obtain $\zeta^{(4)}_{in}=0$, 
while Eq. (\ref{eq:order5}) for $\zeta^{(5)}_{in}$ becomes 
\begin{equation}
{\partial^{2}\zeta_{in}^{(5)}\over \partial\chi^{2}}=
-{a_{0}^{2}\over 8a}\alpha_{0}\chi(\chi^{2}-3\chi_{0}^{2})
-{1\over4}\alpha_{0}a_{0}\chi^{3}. 
\end{equation}
The above equation shows that $\zeta^{(5)}_{in}$ depends on time $\eta$ 
through the first term on the right-hand side and 
hence the assumption of the quasi-static evolution is no longer consistent 
with the solution above $O(\chi_{0}^{5})$.
In the case of $\beta\neq 0$, the quasi-static solution (\ref{eq:qsin}), 
(\ref{eq:qsex})
is also valid up to $O(\chi_0^4)$.
Here we should again note that our approximation based on the smallness of 
$\chi_{0}$ is valid only when $a_{0}/a$ is not so large, i.e., 
$a_{0}/a \ll 1/\chi_{0}$ and further 
$a_{0} /(2a) \ll 1 / \chi_{0}^{3}$.

%************************************%
%*          References              *%
%************************************%

\newpage

%*************************************%
%         Figure Captions             %
%*************************************%

\vskip 0.3in
\centerline{FIGURE CAPTION}
\vskip 0.05in

\newcounter{fignum}
\begin{list}{Fig.\arabic{fignum}.}{\usecounter{fignum}}

%1===============%
\item
Regions of the Oppenheimer-Snyder space-time 
for constant $\theta$ and $\phi$, expressed in characteristic coordinates.
In the ``stationary region'' the initial stationary field configuration
remains a stationary solution.
%2================%
\item
The wave form of the scalar gravitational wave 
in the Brans-Dicke theory at $r=100M$.
The initial dust radius $r_{s0}$ is $10M$.
The ordinate is $\zeta=r\delta\varphi$.
The abscissa is the time $t$ from the beginning of the collapse at
$t=0$. 
%3================%
\item
The scalar field in the interior of the dust 
(the region (A) of Fig. 1) in the Brans-Dicke theory.
The initial dust radius $r_{s0}$ is $10M$.
(a) The ordinate is $\zeta=a\sin\chi\delta\varphi$.
The abscissas are the null coordinates $u=\eta-\chi$ and
$v=\eta+\chi$.
(b) The time evolution of the scalar field seen by comoving observers.
The numbers attached to the curves are the radial coordinates $\chi$
of the comoving observers.
The ordinate is $\zeta=a\sin\chi$ and the abscissa is the conformal time 
$\eta$.
(c) The scalar field configuration on the spacelike hypersurface of the 
conformal time $\eta=const.$.
The abscissa is the radial coordinate
$\chi$. The numbers attached to the curves are the value of $\eta$.
The curve with the label ``QS'' shows the approximate quasi-static
solution obtained in the text.
See the text for further details.
%4================%
\item
The scalar field in the exterior of the dust 
(the region (B) of Fig. 1) in the Brans-Dicke theory.
The initial dust radius $r_{s0}$ is 10M.
The ordinate is $\zeta=r\delta\varphi$.
The abscissas are the null coordinates $u$ and $v$ extended 
from the interior region to the exterior.
%5================%
\item
The scalar field in the exterior of the dust 
(the region (C) of Fig. 1) in the Brans-Dicke theory.
The initial dust radius $r_{s0}$ is 10M.
The ordinate is $\zeta=r\delta\varphi$.
The abscissas are the null coordinates $\tilde{u}=t-r_*$ and
$\tilde{v}=t+r_*$.
%6================%
\item
The wave forms of 
the scalar gravitational waves 
from the collapses of various initial radii 
in the Brans-Dicke theory.
The ordinate and abscissa are the same as Fig. 2.
The numbers attached the curves represent the value of the initial radii
respectively.
%7================%
\item
Relation between the peak amplitude 
and initial dust radius $r_{s0}$
in the Brans-Dicke theory.
%8================%
\item
Relation between the characteristic frequency $f_c$
and initial dust radius $r_{s0}$
in the Brans-Dicke theory.
%9================%
\item
Relation between the time scale of the last bounce and 
initial dust radius $r_{s0}$.
The ordinate is the time scale of the last oscillation
which is defined as the half-value width of that bounce.
%10===============%
\item
The wave form of the scalar gravitational wave 
for various $\beta$ at $r=100M$.
The numbers attached to the curves represent the value of $\beta$
respectively. The initial dust radius $r_{s0}$ is fixed to $10M$.
%11===============%
\item
Same as Fig. 10, but for different $\beta$.
%12===============%
\item
Same as Fig. 10, but for different $\beta$.
%13===============%
\item
The time of the first minimum of the wave form
from the reach of the first ray as a function of $\beta>0$.
%14===============%
\item
The time evolution 
of the scalar field seen by comoving observers 
in the interior of the dust (the region (A) of Fig. 1)
for $\beta=50$ .
The ordinate and abscissa are same as Fig. 3(b).
%15===============%
\item
The time evolution 
of the scalar field seen by comving observers 
in the interior of the dust (the region (A) of Fig. 1)
for $\beta=-5$ .
The ordinate and abscissa are same as Fig. 3(b).

\end{list}

\newpage

\begin{figure}
   \begin{center}
      \begin{picture}(200,200)
         \put(15,15){\line(1,0){165}}
         \put(15,15){\line(0,1){70}}
         \put(65,15){\line(1,1){135}}
         \put(15,85){\line(1,1){115}}
         \put(65,15){\line(0,1){120}}
         \put(65,135){\line(1,-1){60}}
         \put(15,85){\line(1,1){100}}
         \put(15,0){ $\eta=0 $ }
         \put(65,0){initial hypersurface}
         \put(125,20){ $t=0$ }
         \put(0,40){$\chi=0$}
         \put(130,50){stationary region}
         \put(160,135){first ray $\tilde{u}=\tilde{u_0}$}
         \put(60,160){horizon}
         \put(80,100){$\tilde{v}=\tilde{v_0}$}
         \put(10,140){$u=\pi-3\chi_0$}
         \put(30,70){(A)}
         \put(80,70){(B)}
         \put(125,120){(C)}
       \end{picture}
   \end{center}
   \begin{center}
Fig. 1
   \end{center}
\end{figure}

\begin{figure}
      \vspace{1cm}
      \centerline{\epsfysize 9cm \epsfxsize 13cm \epsfbox{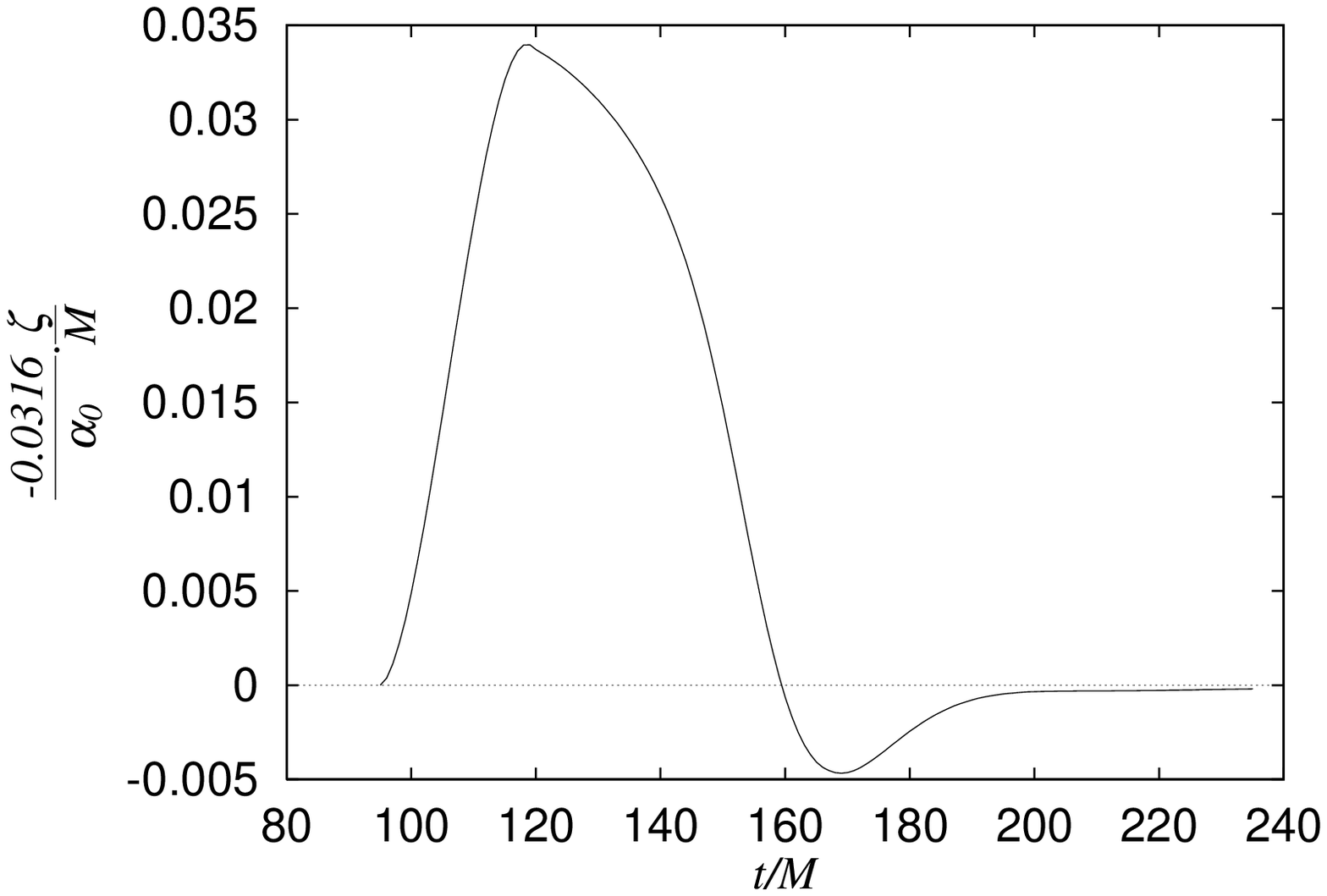}}
   \begin{center}
Fig. 2
   \end{center}
\end{figure}

\begin{figure}
      \vspace{1cm}
      \centerline{\epsfysize 20cm \epsfxsize 13cm \epsfbox{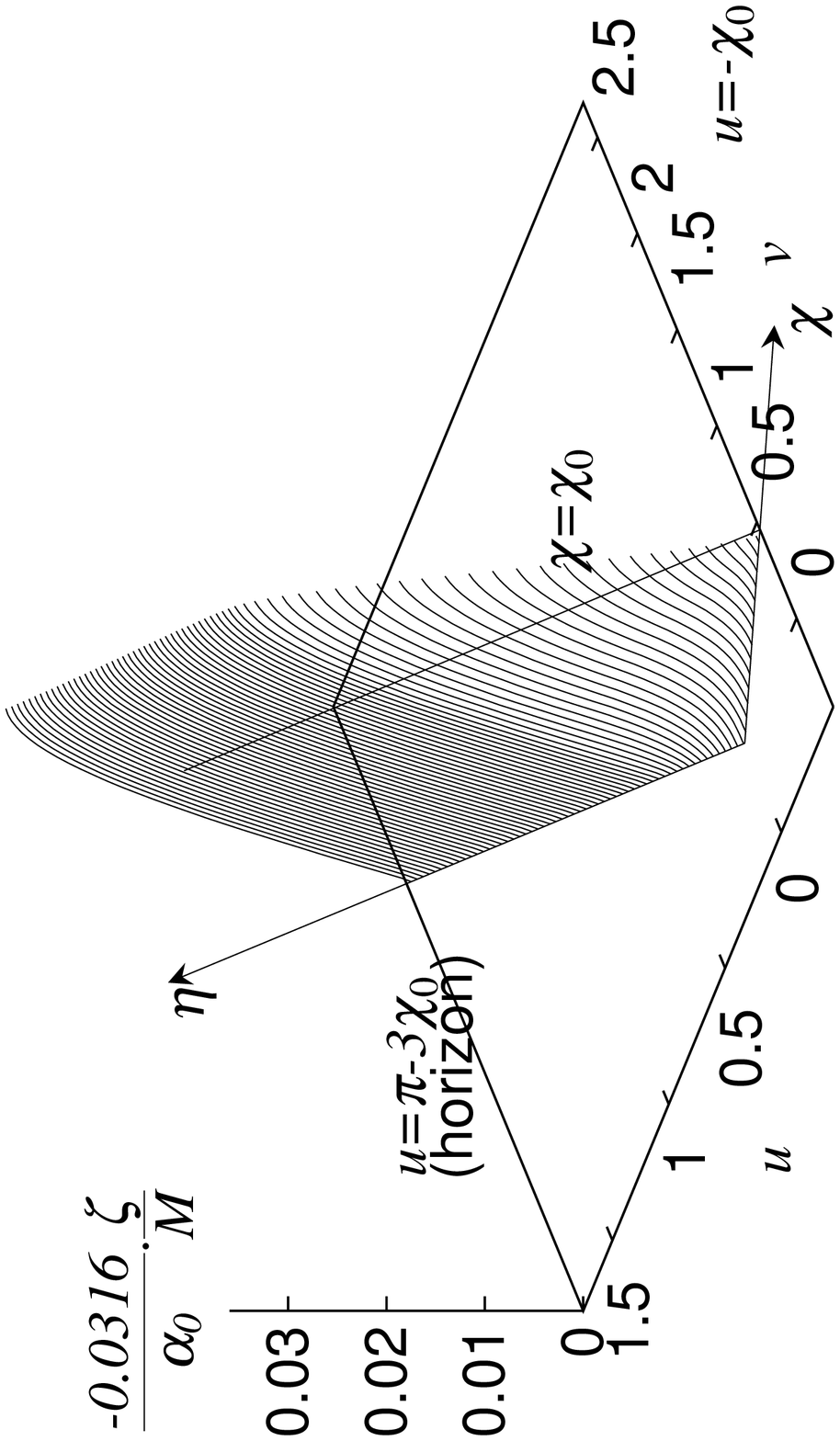}}
   \begin{center}
Fig. 3(a)
   \end{center}
\end{figure}

\begin{figure}
      \vspace{1cm}
      \centerline{\epsfysize 9cm \epsfxsize 13cm \epsfbox{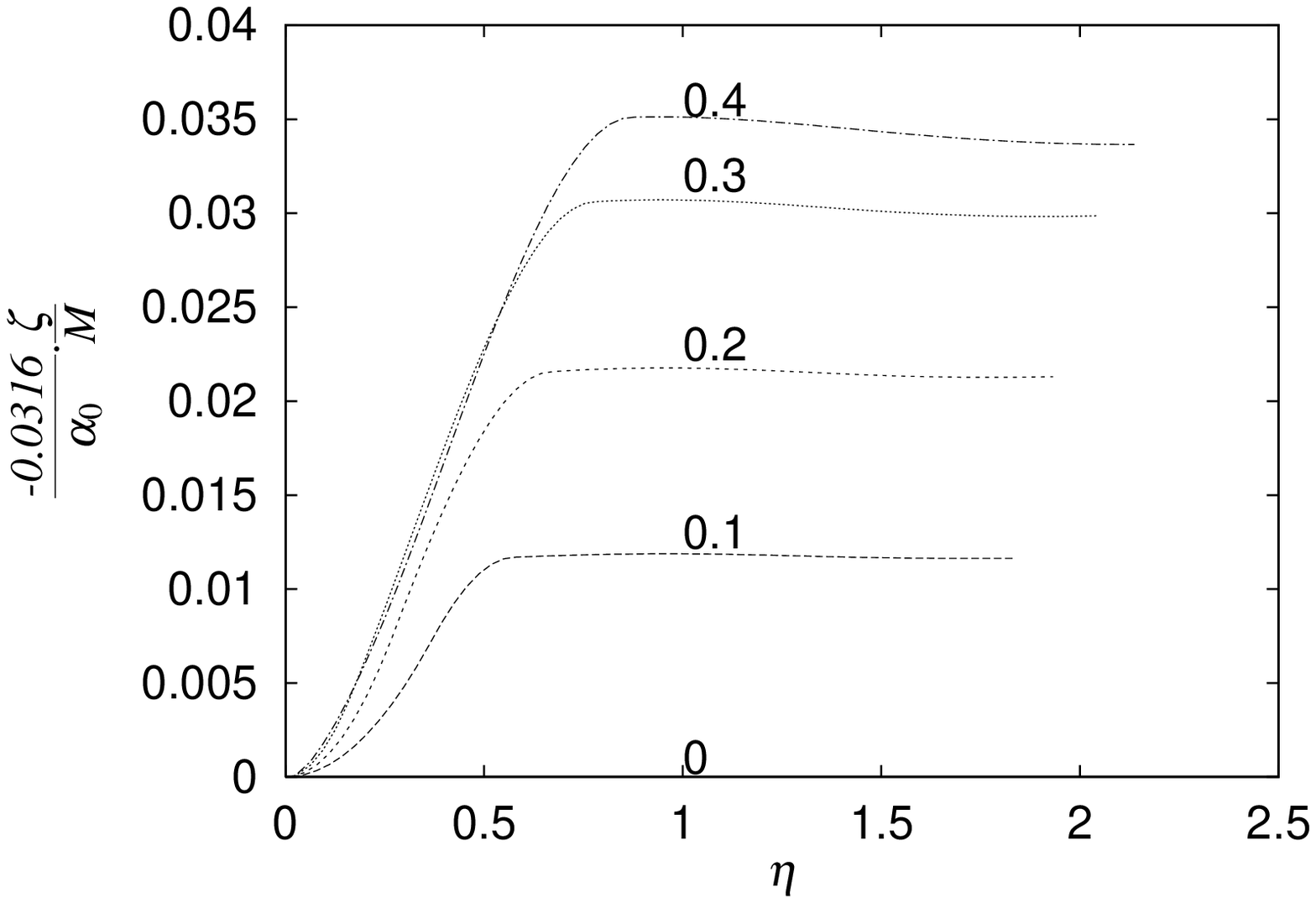}}
   \begin{center}
Fig. 3(b)
   \end{center}
\end{figure}

\begin{figure}
      \vspace{1cm}
      \centerline{\epsfysize 9cm \epsfxsize 13cm \epsfbox{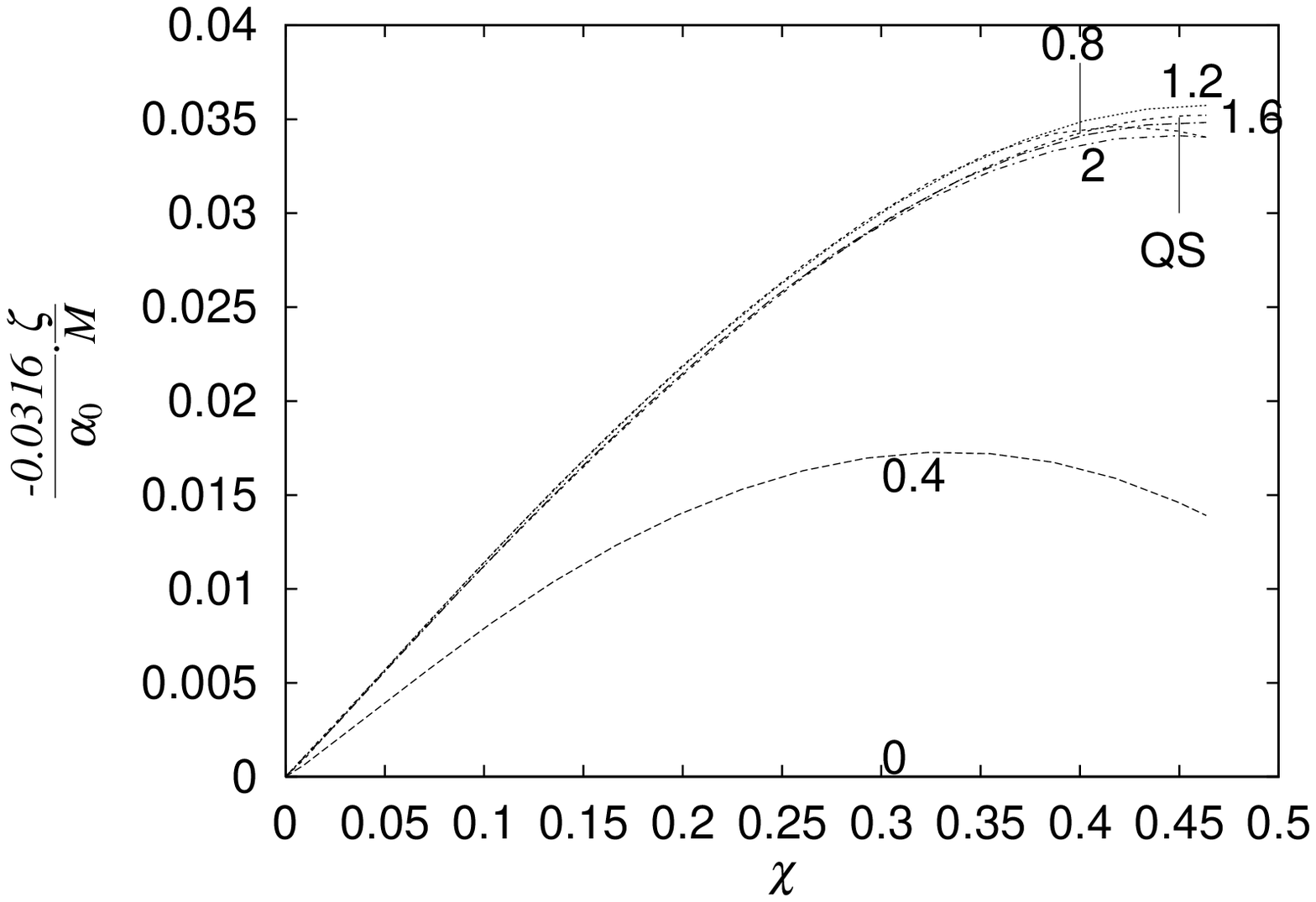}}
   \begin{center}
Fig. 3(c)
   \end{center}
\end{figure}

\begin{figure}
      \vspace{1cm}
      \centerline{\epsfysize 20cm \epsfxsize 13cm \epsfbox{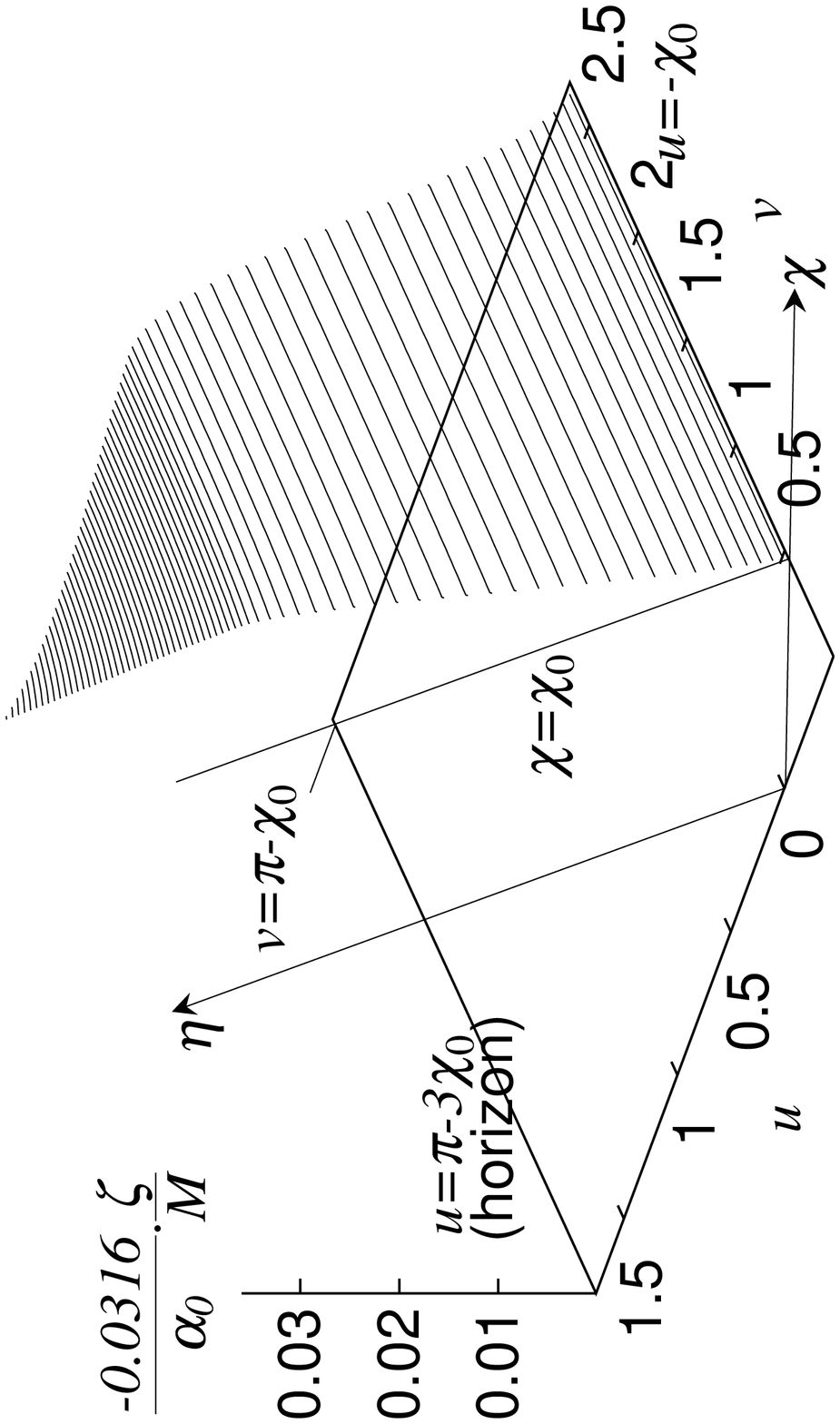}}
   \begin{center}
Fig. 4
   \end{center}
\end{figure}

\begin{figure}
      \vspace{1cm}
      \centerline{\epsfysize 20cm \epsfxsize 13cm \epsfbox{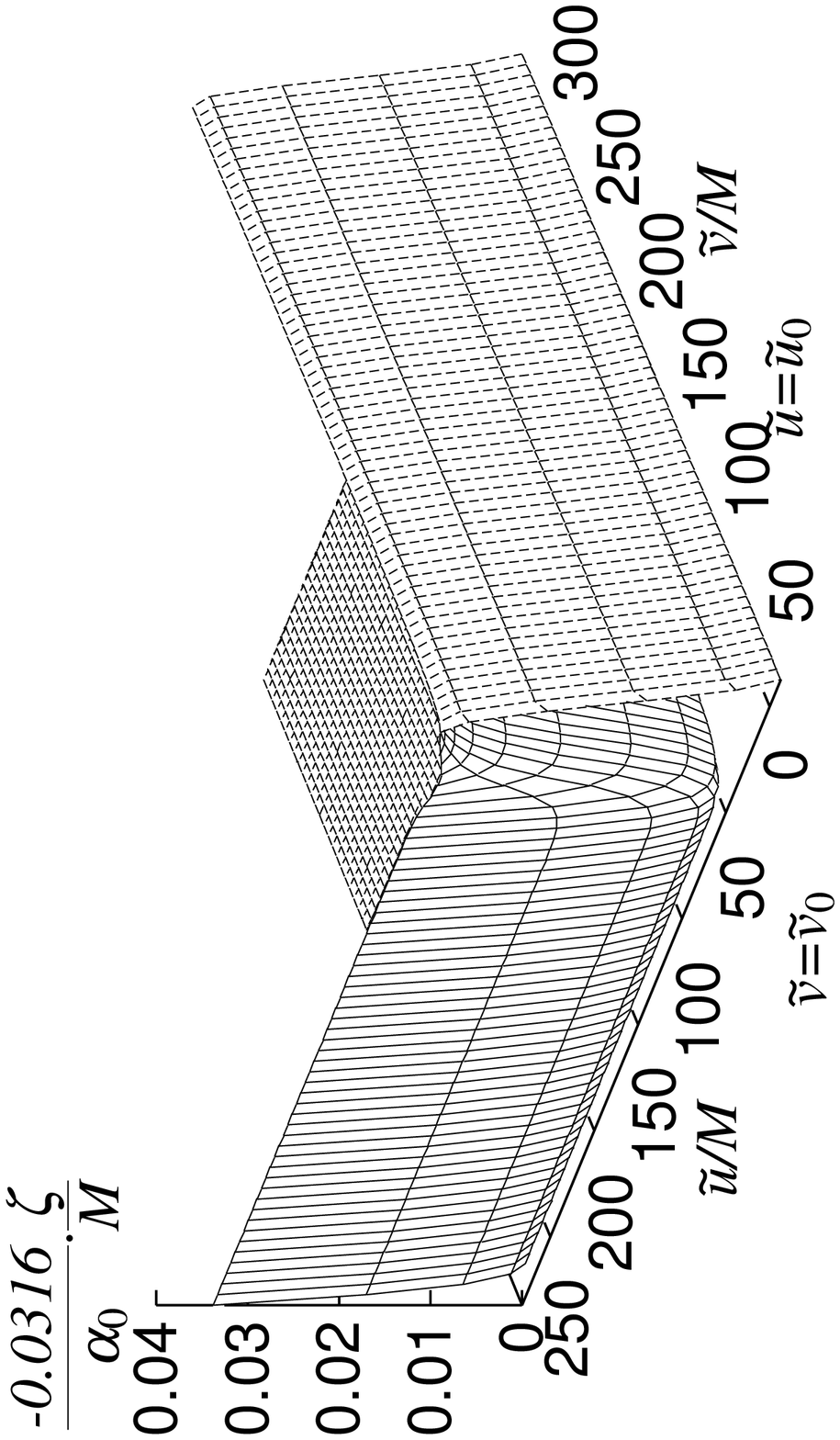}}
   \begin{center}
Fig. 5
   \end{center}
\end{figure}

\begin{figure}
      \vspace{1cm}
      \centerline{\epsfysize 9cm \epsfxsize 13cm \epsfbox{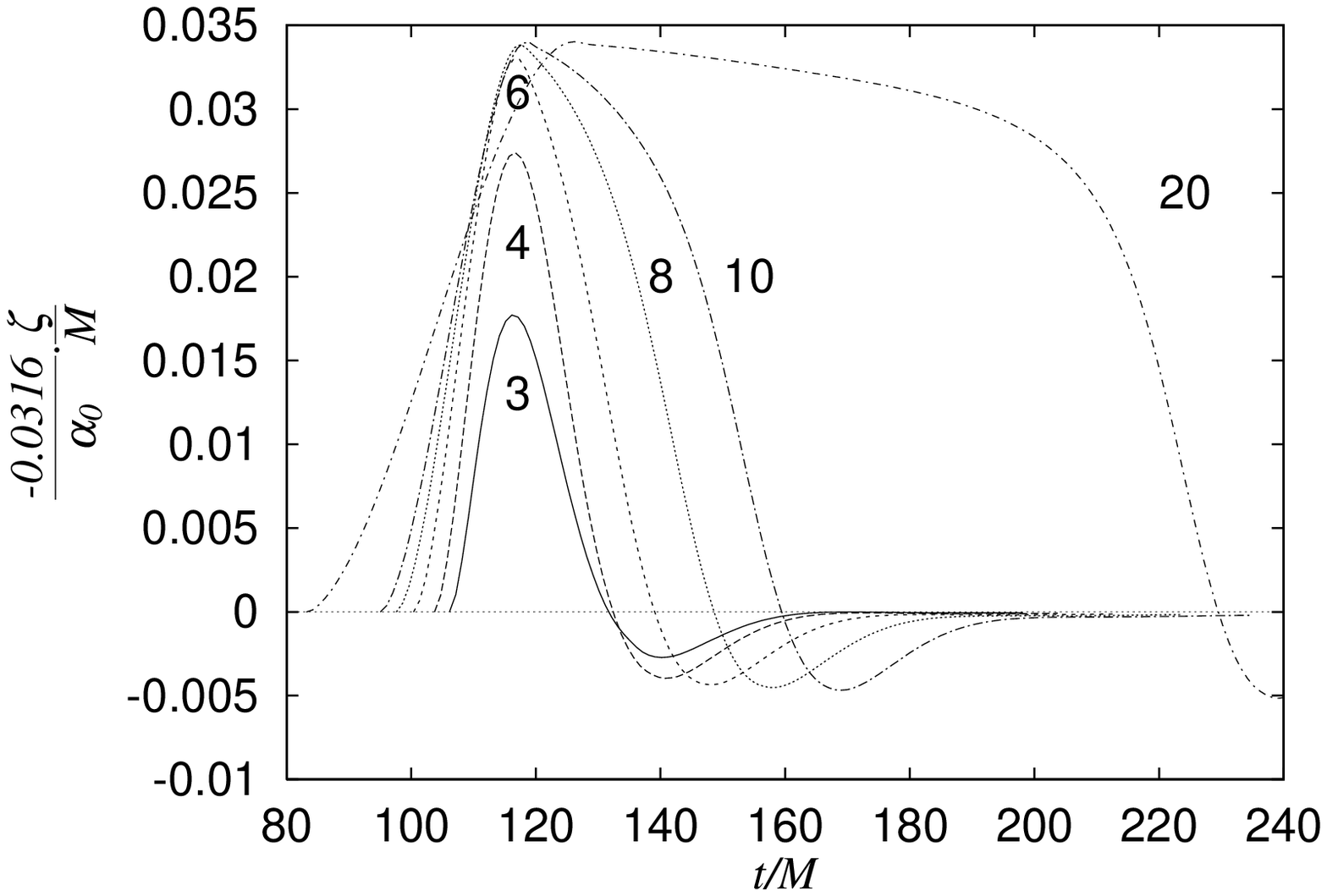}}
   \begin{center}
Fig. 6
   \end{center}
\end{figure}

\begin{figure}
      \vspace{1cm}
      \centerline{\epsfysize 9cm \epsfxsize 13cm \epsfbox{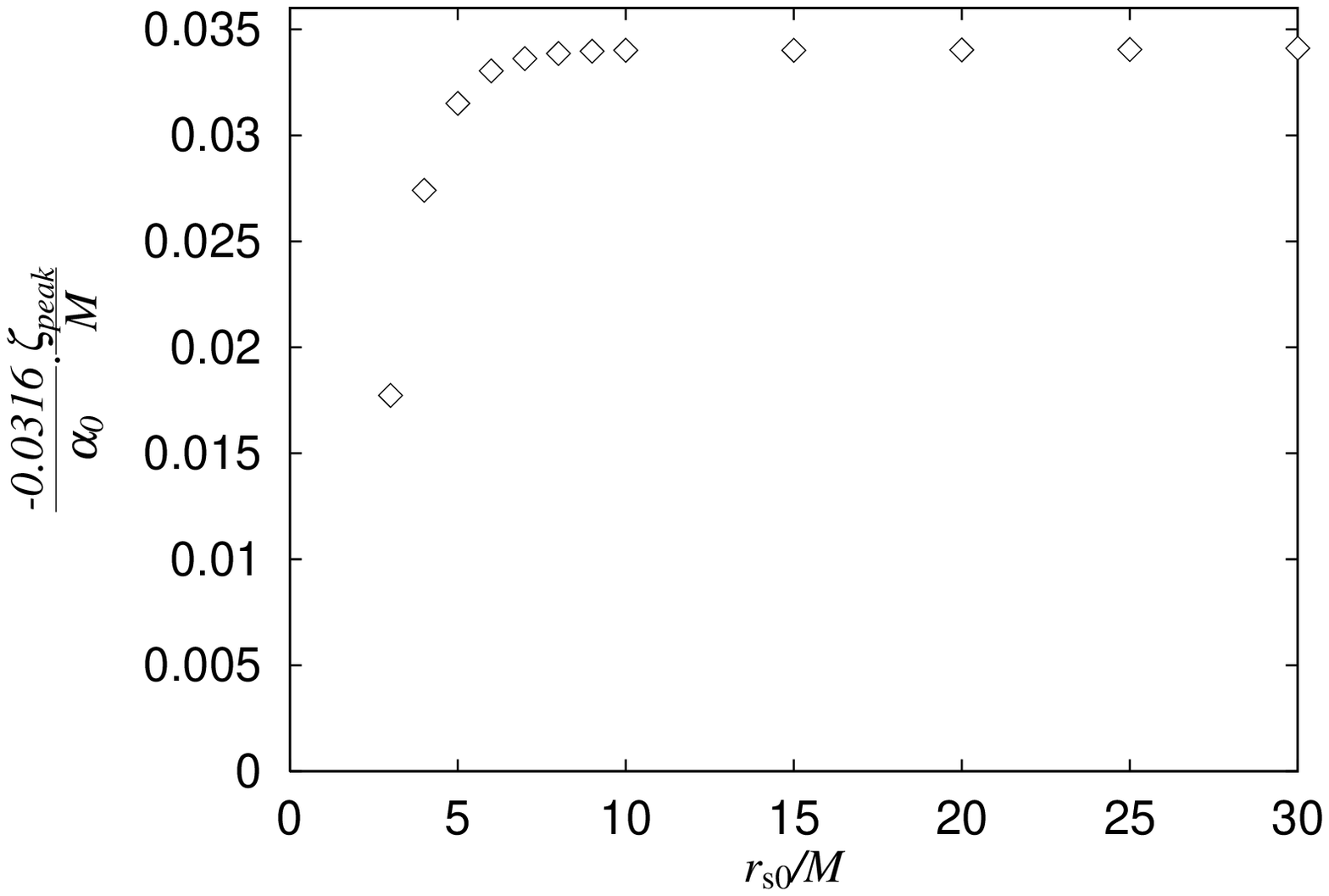}}
   \begin{center}
Fig. 7
   \end{center}
\end{figure}

\begin{figure}
      \vspace{1cm}
      \centerline{\epsfysize 9cm \epsfxsize 13cm \epsfbox{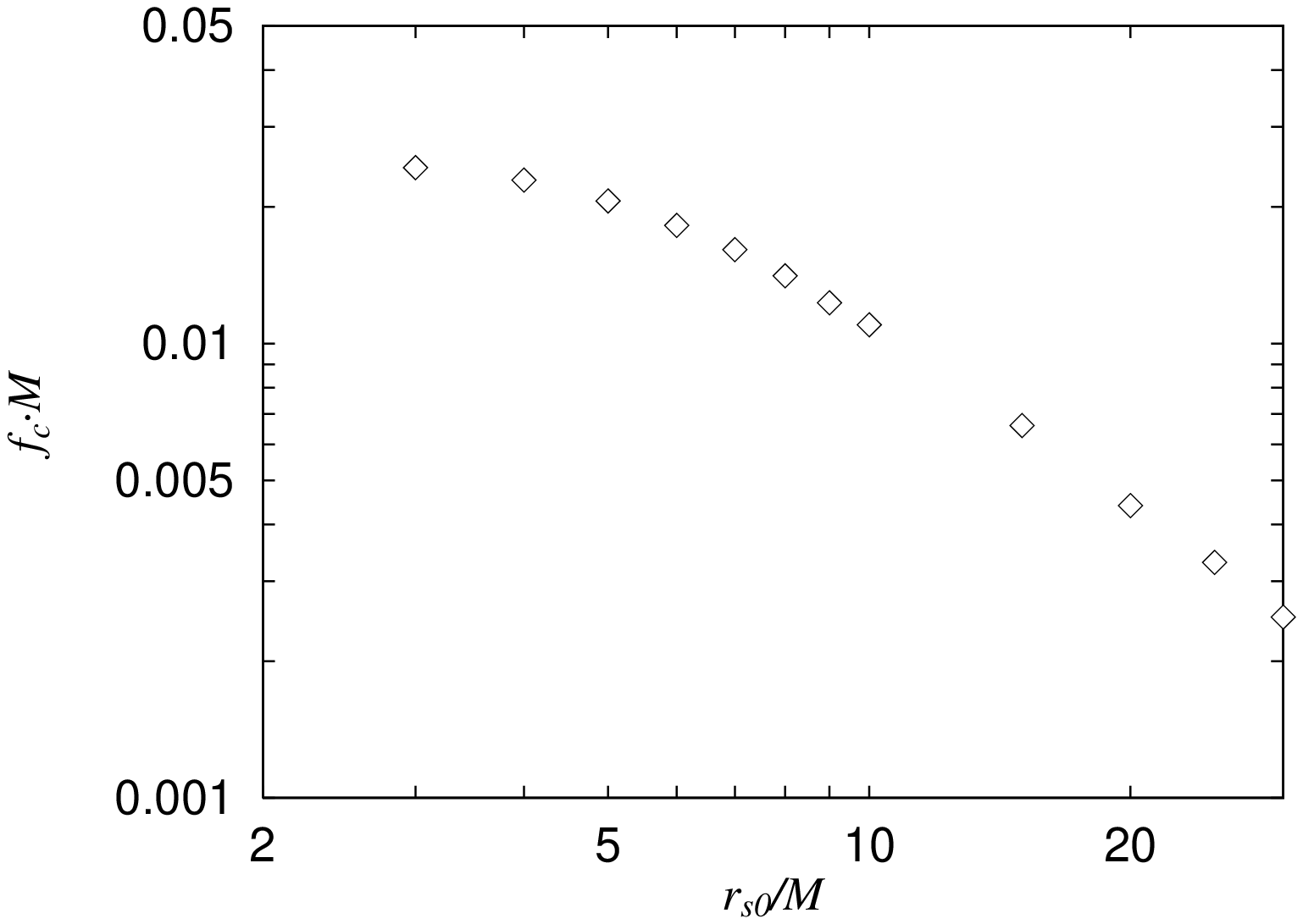}}
   \begin{center}
Fig. 8
   \end{center}
\end{figure}

\begin{figure}
      \vspace{1cm}
      \centerline{\epsfysize 9cm \epsfxsize 13cm \epsfbox{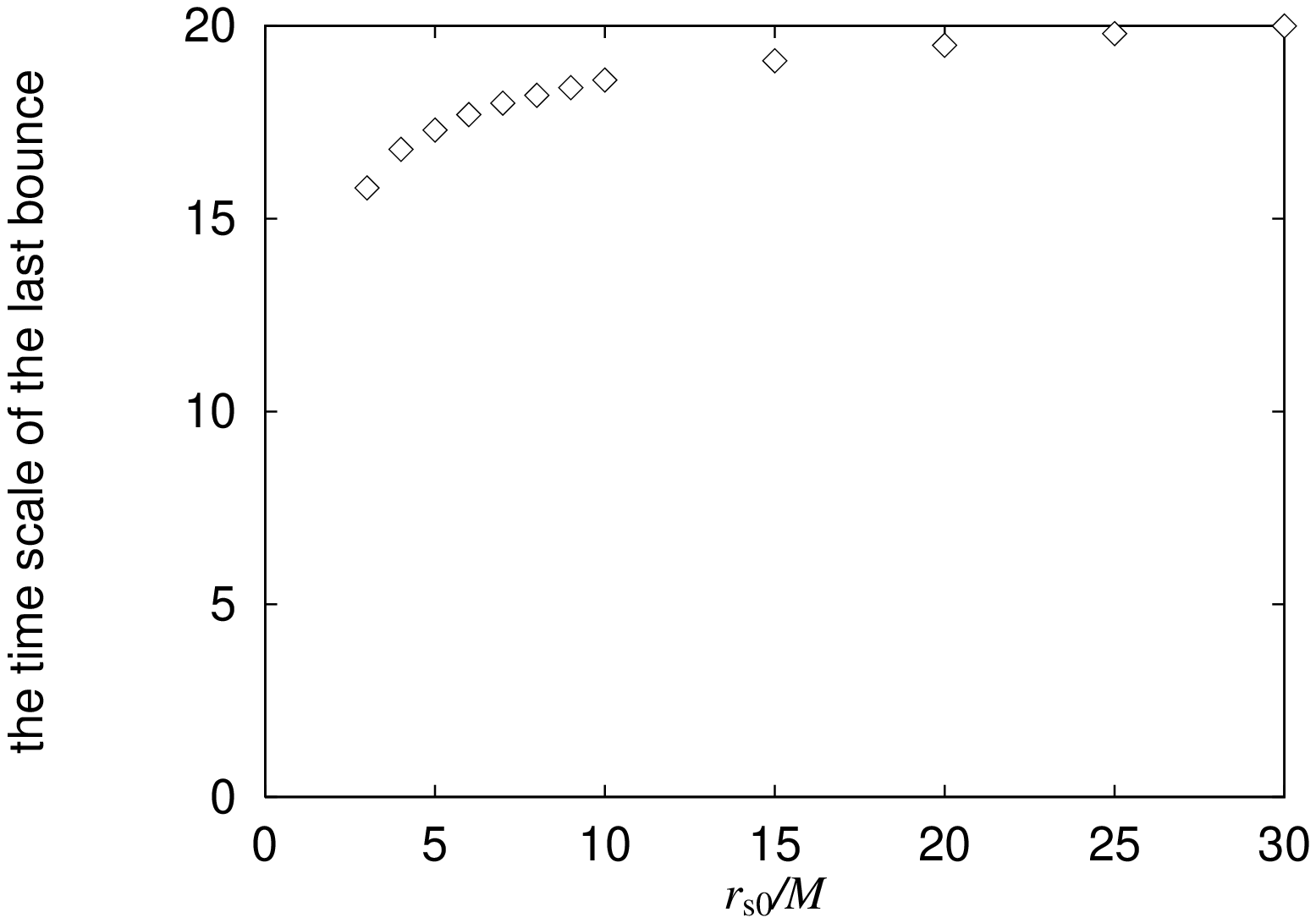}}
   \begin{center}
Fig. 9
   \end{center}
\end{figure}

\begin{figure}
      \vspace{1cm}
      \centerline{\epsfysize 9cm \epsfxsize 13cm \epsfbox{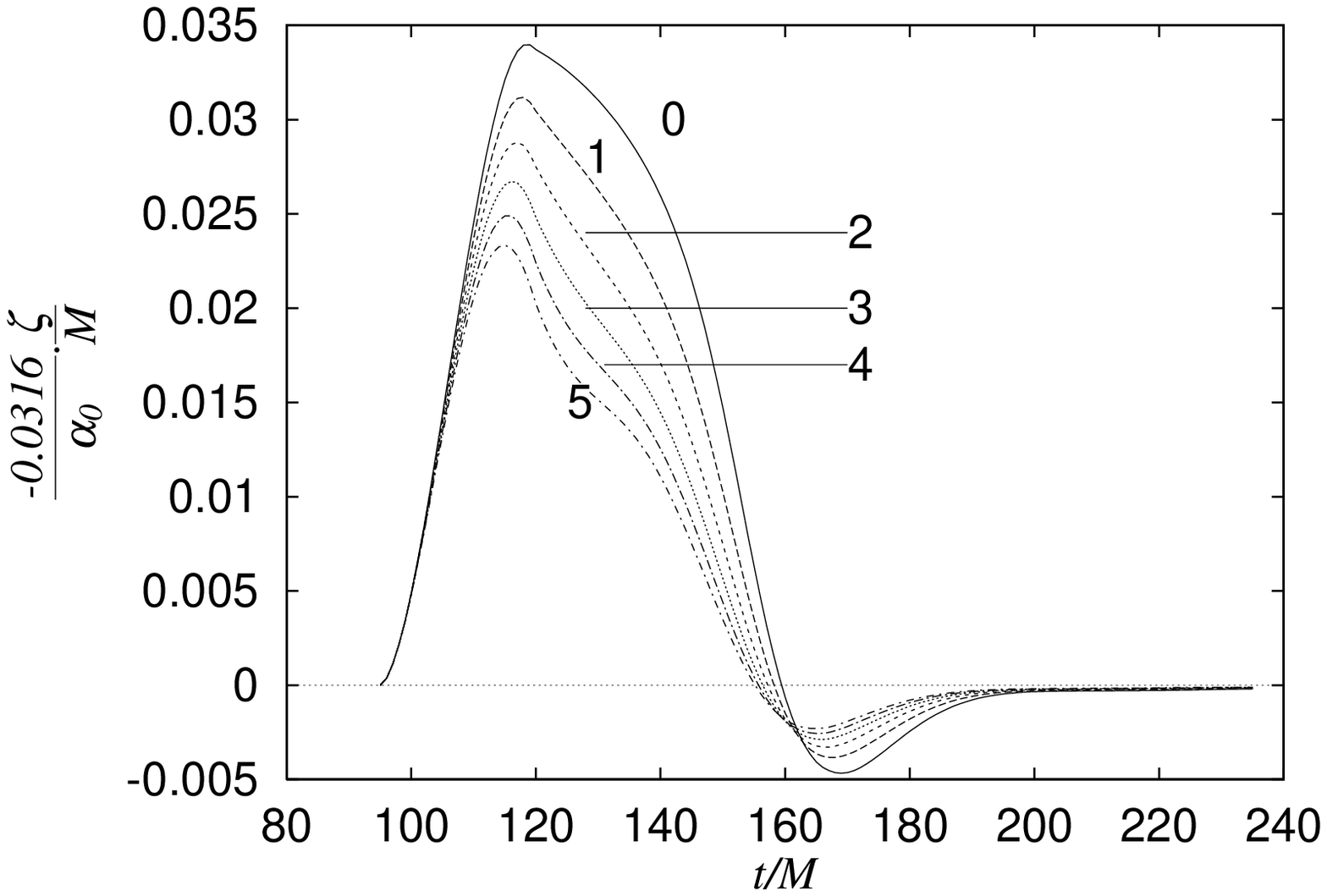}}
   \begin{center}
Fig. 10
   \end{center}
\end{figure}

\begin{figure}
      \vspace{1cm}
      \centerline{\epsfysize 9cm \epsfxsize 13cm \epsfbox{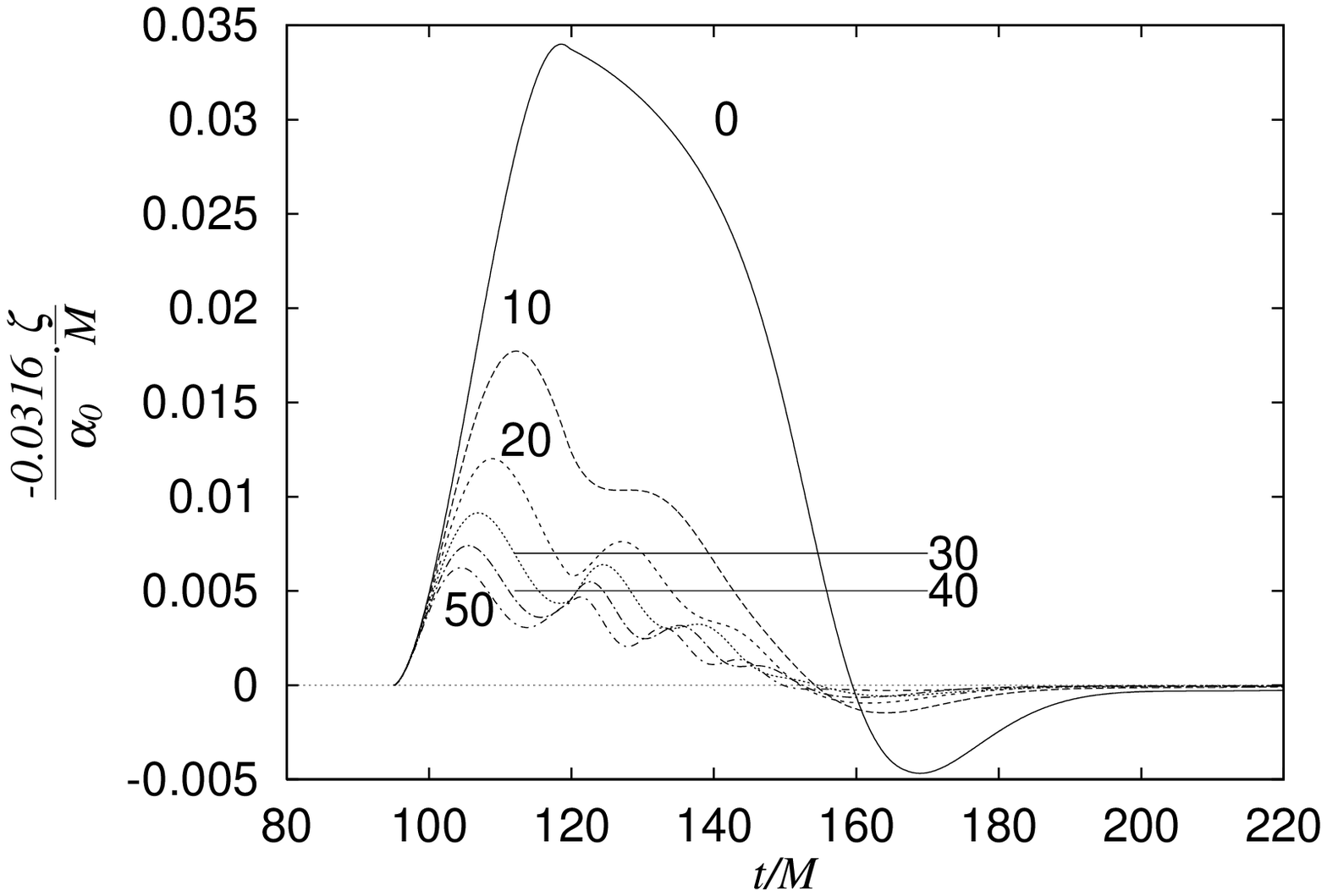}}
   \begin{center}
Fig. 11
   \end{center}
\end{figure}

\begin{figure}
      \vspace{1cm}
      \centerline{\epsfysize 9cm \epsfxsize 13cm \epsfbox{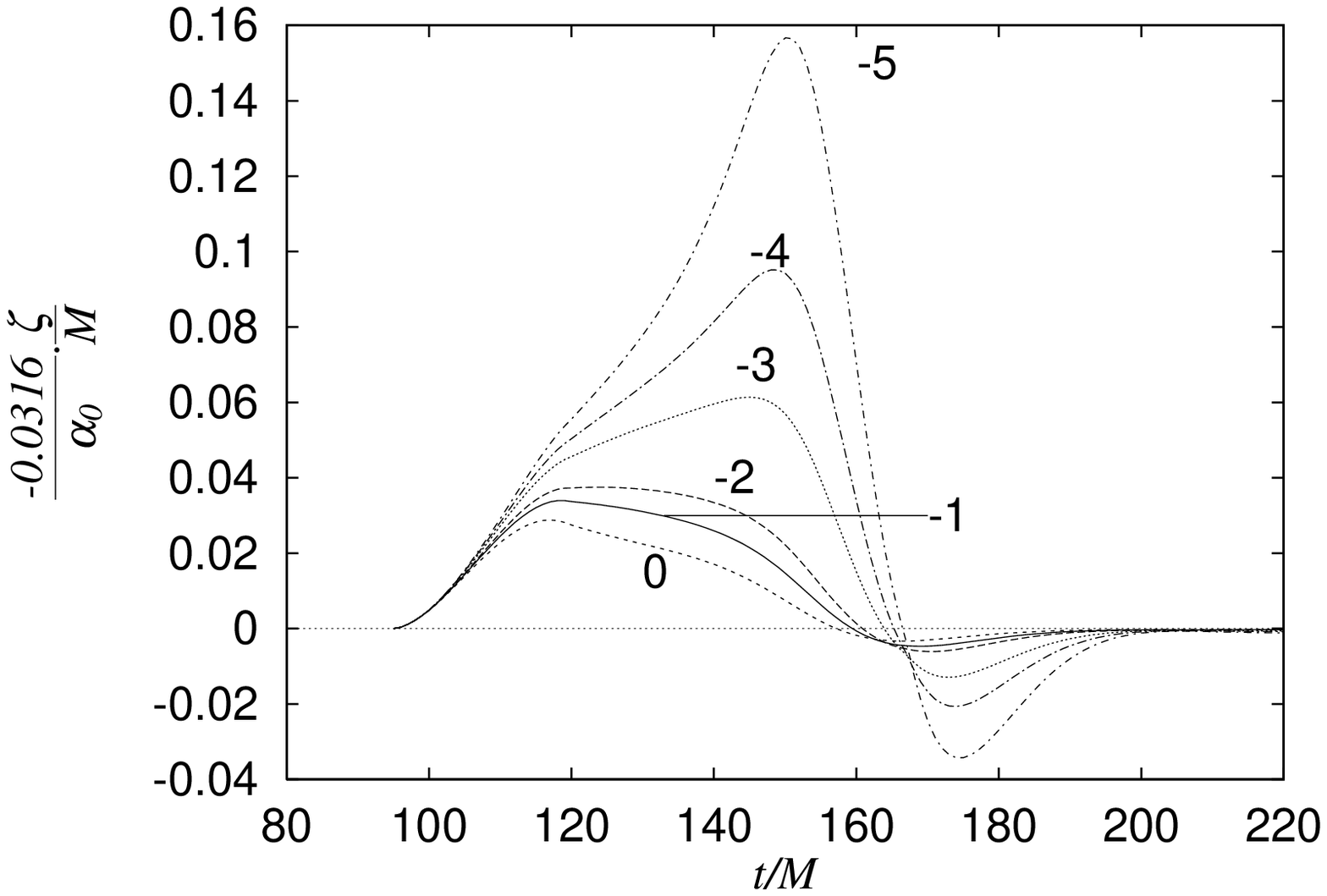}}
   \begin{center}
Fig. 12
   \end{center}
\end{figure}
      
\begin{figure}
      \vspace{1cm}
      \centerline{\epsfysize 9cm \epsfxsize 13cm \epsfbox{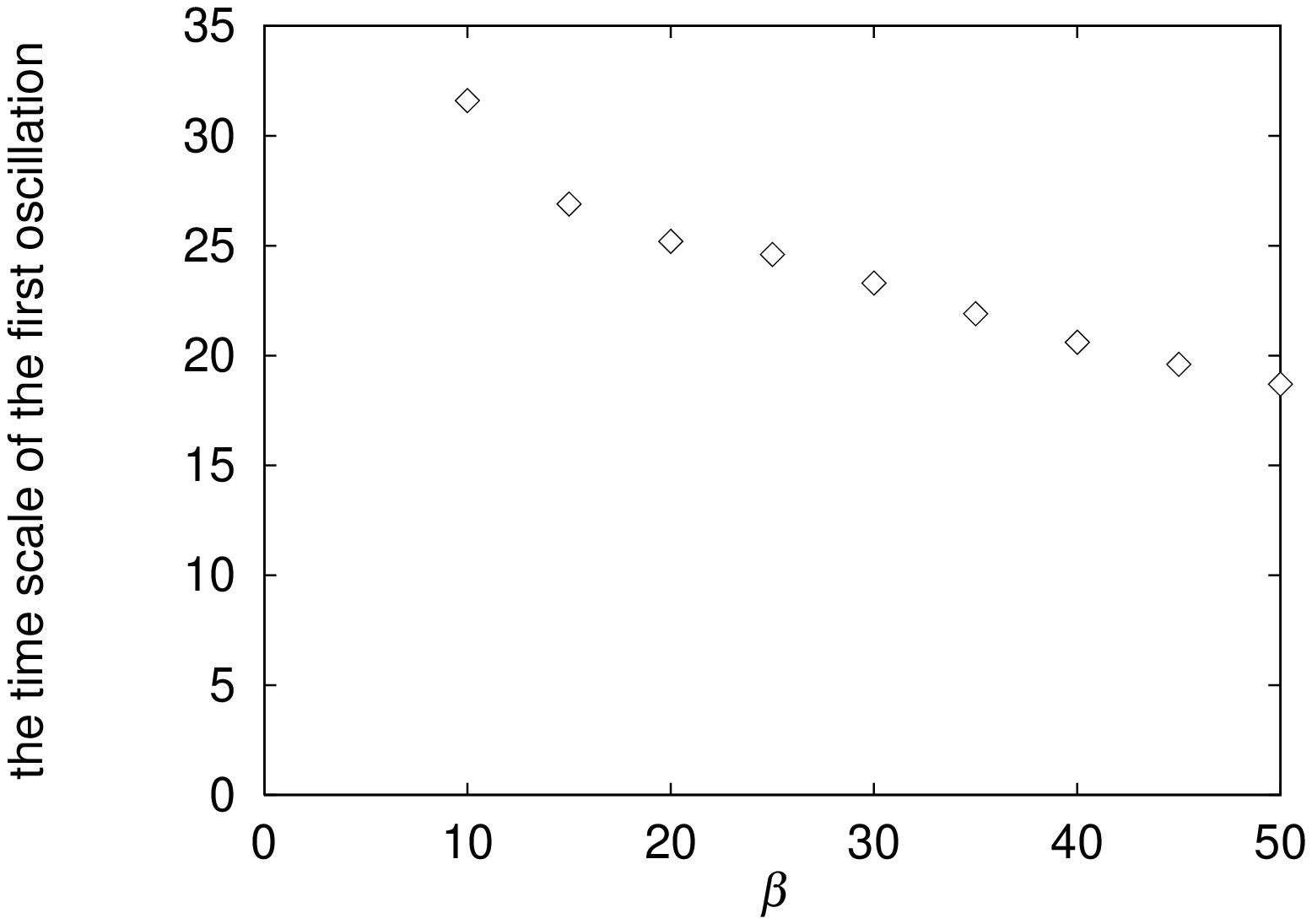}}
   \begin{center}
Fig. 13
   \end{center}
\end{figure}
      
\begin{figure}
      \vspace{1cm}
      \centerline{\epsfysize 9cm \epsfxsize 13cm \epsfbox{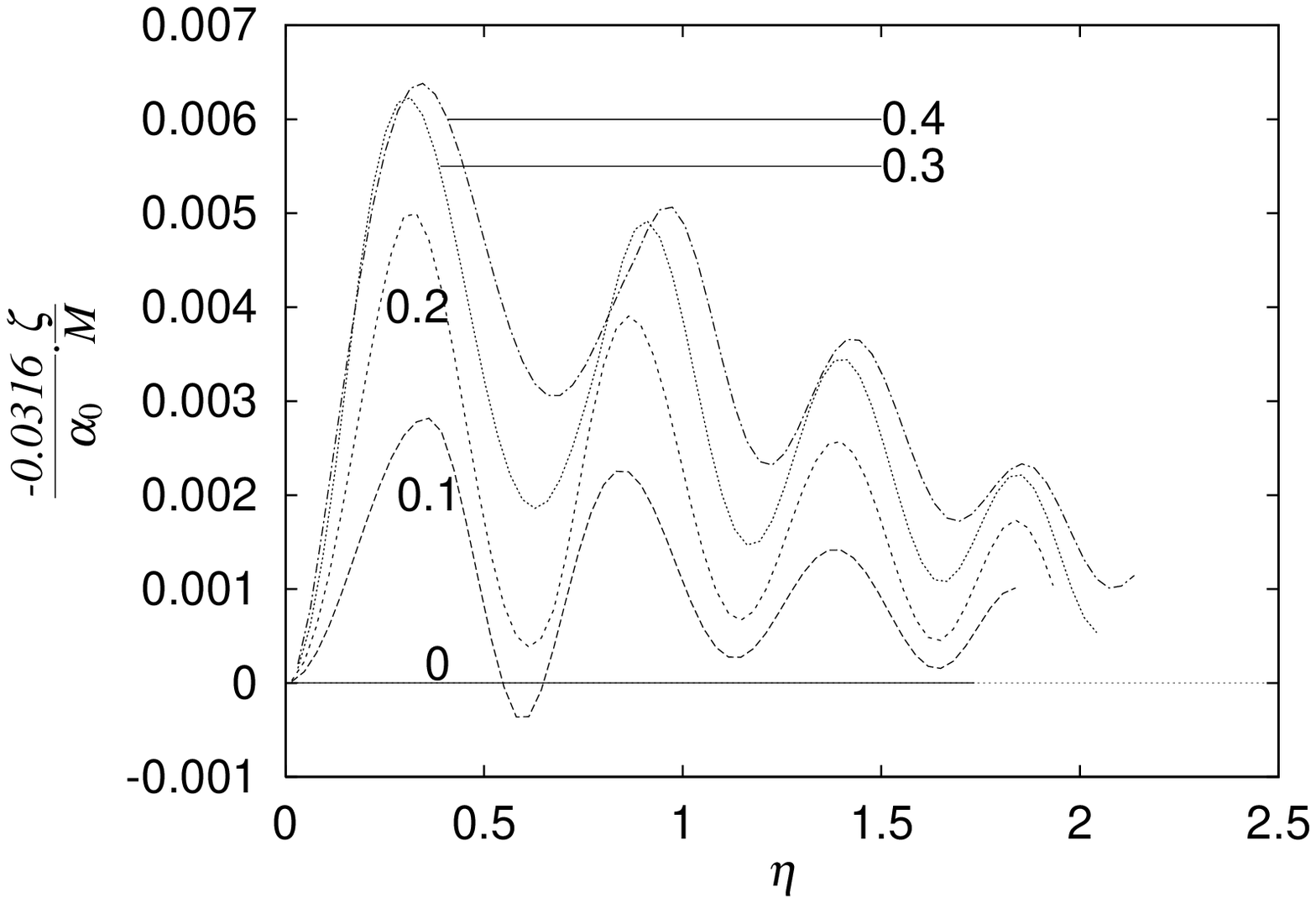}}
   \begin{center}
Fig. 14
   \end{center}
\end{figure}
      
\begin{figure}
      \vspace{1cm}
      \centerline{\epsfysize 9cm \epsfxsize 13cm \epsfbox{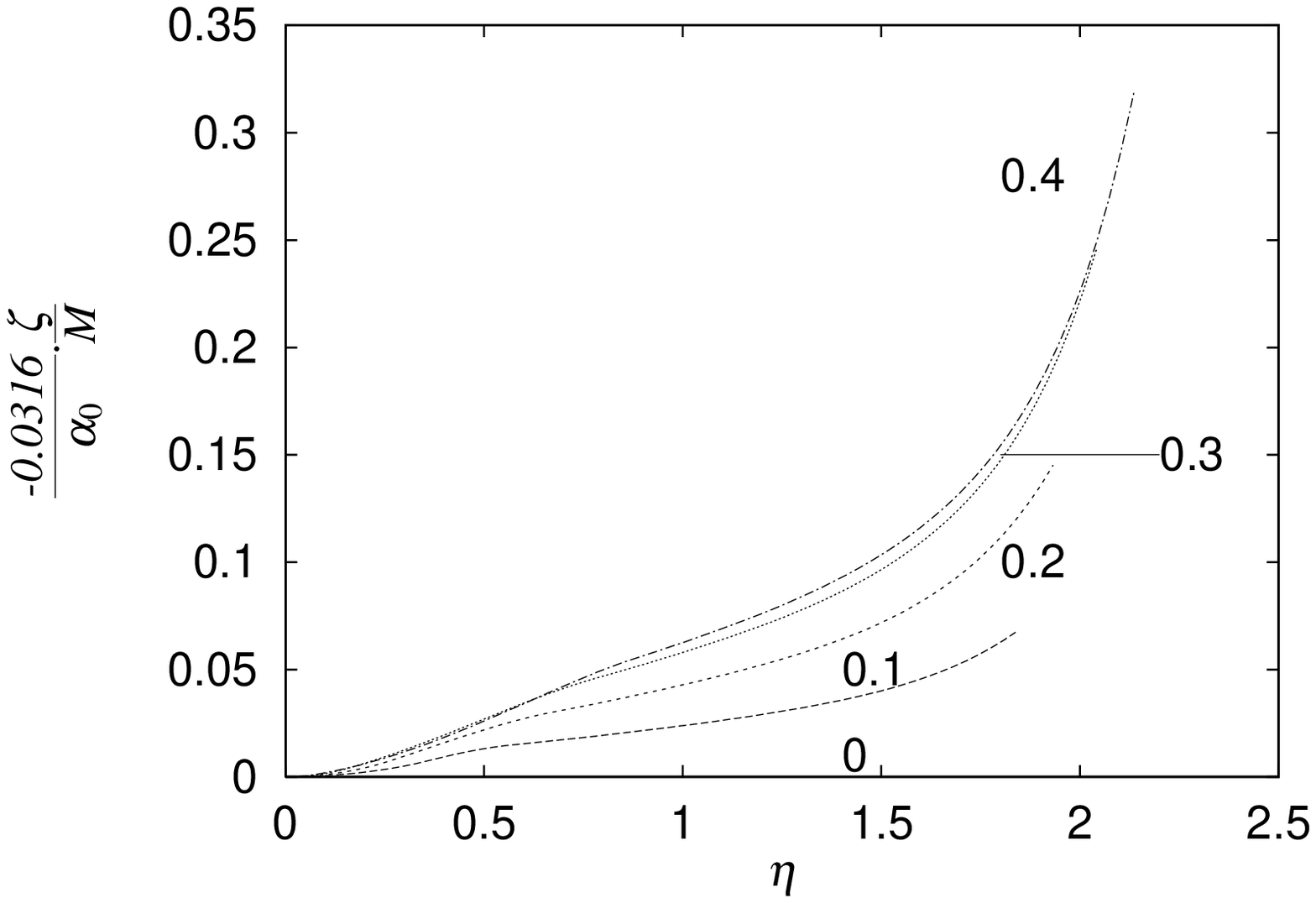}}
   \begin{center}
Fig. 15      
   \end{center}
    \end{figure}

\end{document}